\documentclass[journal]{IEEEtran}
\ifCLASSINFOpdf
\else
\fi
\usepackage[cmex10]{amsmath}
\usepackage{graphicx}
\usepackage{subfigure}
\usepackage{multirow}
\usepackage{soul}
\usepackage{amsmath}
\usepackage{url}
\usepackage{algorithm}
\usepackage[noend]{algpseudocode}
\usepackage{adjustbox}
\usepackage{amssymb}
\usepackage{setspace}

\usepackage[cmex10]{amsmath}
\usepackage{graphicx}
\usepackage{subfigure}
\usepackage{multirow}
\usepackage{soul}
\usepackage{amsmath}
\usepackage{url}
\usepackage{adjustbox}
\usepackage[table,xcdraw]{xcolor}
\usepackage[roman]{parnotes}
\usepackage[roman]{parnotes}
\usepackage{amssymb}

\usepackage{rotating}
\usepackage{tikz}

\soulregister\cite7
\soulregister\ref7
\soulregister\pageref7

\begin{document}
%
\title{\fontsize{16}{24}\selectfont Dependability Analysis of Data Storage Systems in Presence of Soft Errors}
%
%
%

\author{Mostafa~Kishani,
        Mehdi~Tahoori,~\IEEEmembership{Senior Member,~IEEE,}
        and~Hossein~Asadi,~\IEEEmembership{Senior Member,~IEEE}
\thanks{Mostafa Kishani is a PhD graduate from Department of Computer Engineering, Sharif University of Technology, Tehran, Iran (e-mail: kishani@ce.sharif.edu).}
\thanks{Mehdi Tahoori is a full professor and Chair of Dependable Nano-Computing (CDNC) at the Institute of Computer Science \& Engineering (ITEC), Department of Computer Science, Karlsruhe Institute of Technology (KIT), Germany. (e-mail: mehdi.tahoori@kit.edu) }
\thanks{Hossein Asadi is an Associate Professor in Department of Computer Engineering, Sharif University of Technology, Tehran, Iran (e-mail: asadi@sharif.edu).}}

%
%

{}
%



\maketitle

\begin{abstract}
In recent years, high availability and reliability of \emph{Data Storage Systems} (DSS) have been significantly threatened by soft errors occurring in storage controllers. 
Due to their specific functionality and hardware-software stack, error propagation and manifestation in DSS is quite different from general-purpose computing architectures. 
To our knowledge, no previous study has examined the system-level effects of soft errors on the availability and reliability of data storage systems.
In this paper, we first analyze the effects of soft errors occurring in
the server processors of storage controllers on the entire storage system
dependability. 
To this end, we implemented the major functions
of a typical data storage system controller, running on a full
stack of storage system operating system, and developed a framework to perform
fault injection experiments using a full system simulator. 
We then propose a new metric, \emph{Storage System Vulnerability Factor} (SSVF), to accurately capture the impact of soft errors in storage systems.
{By conducting extensive experiments, it is revealed that depending
on the controller configuration, up to 40\% of cache memory
contains end-user data where any unrecoverable soft errors in
this part will result in \emph{Data Loss} (DL) in an irreversible manner.}
However, soft errors in the rest of cache memory filled by \emph{Operating System} (OS) and storage applications will
result in \emph{Data Unavailability} (DU) at the storage system level. 
Our analysis also shows that \emph{Detectable Unrecoverable Errors} (DUEs) on the cache data field are the major cause of DU in storage systems,
while \emph{Silent Data Corruptions} (SDCs) in the cache tag and data field are mainly the cause of DL in storage systems.
\end{abstract}


%
\IEEEpeerreviewmaketitle

\section{Introduction}
\label{sec:Intro}
The increasing demands of data intensive applications 
have made IT infrastructures 
mainly rely on \emph{Data Storage Systems} (DSS), which tend to assure the storage of data with high dependability and performance\footnote{
High dependability and performance of data storage system is achieved by mechanisms such as system-level and component-level redundancy, caching~\cite{salkhordeh2018,ahmadian2018eci}, and data tiering~\cite{salkhordeh2015}.
}. 
{According to \emph{International Data Corporation} (IDC) Worldwide Quarterly Enterprise Storage System Tracker, the worldwide enterprise storage system revenue reached \$10.8 billion in the second quarter of 2017~\cite{idcstoragemarket}, while another report by \emph{MarketsandMarkets} forecasts storage market is valued at \$144.76 billion by 2022~\cite{marketsandmarkets}.}  
Meanwhile, the cost of downtime and data loss is the most critical challenge of storage systems~\cite{kishani-tr-2018,kishani2017}.
A survey by CA Technologies shows that North American businesses are annually losing \$26.5 billion in revenue through downtime and data recovery, 
while the average loss of each company is \$160,000 per year~\cite{acodreport}.    
Another survey reports \$273 million downtime cost in 2007-2013 for 28 Cloud service providers~\cite{gagnaire2012downtime}. 

The architecture of DSS has been evolved in time, while it can be roughly categorized into three generations~\cite{storagegeneration-2014}. 
The first generation, \emph{Monolithic Storage Systems}, is attributed by its centralized storage controller, custom designed hardware for controllers, and embedded software, providing a high level of reliability for demanding environments {at} a huge design and manufacturing cost which makes its market very 
limited.   
The second generation, \emph{Modular Storage Systems}, typically use active-active dual controller, providing redundant access to attached storage devices
via independent host interfaces. 
{Due to employing off-the-shelf hardware/software components, these systems have a reduced cost, making them the most popular choice for enterprise markets. 
This architecture is also popular in mid-tier markets when high dependability is demanded.}
Due to its dominance in the storage market, this architecture has been our reference in this study.
The third generation, \emph{Scale-out Storage Systems}, is named after using a cluster of independent, networked, storage nodes. 
Each node can have an architecture similar to the modular systems, with single or dual controller, dedicated cache, and dedicated or shared storage.
Regardless of the generation of storage systems, the end-user data may reside in 
the controller local cache memory, DSS \emph{Global Memory} (GM), or storage (disk) subsystem~\cite{gnanasundaram2012information,chang2008bigtable,nishtala2013scaling}. 
Due to such architectures, unlike general purpose systems,  
failures in the server processor of storage controllers can result in the irreversible loss of end-user data.

Field studies show that among all root causes of storage failures, including software errors and hardware malfunctions (in disks, interconnects, processors, power system~\cite{ahmadian2018ssdRel}, and cooling system), soft errors in the server processors integrated in the storage controllers have a considerable contribution in the storage failures~\cite{shazli-TEST-2008,jiang-ACM_TOS-2008}.
Soft errors, also known as \emph{Single Event Upsets} (SEUs), are transient errors in the memory cells (such as SRAMs) and combinational logic, caused by 
cosmic rays and alpha particles from impurities in packaging materials~\cite{zhang2005robust,lantz1996soft,dasgupta1991material}.
The occurrence of soft errors in memory systems can result in a single bit flip or multiple bit flips, called \emph{Multiple Bit Upset} (MBU). 
Shazli et al. show that more than 15\% of drastic processor failures in storage controllers are caused by soft errors initiated by SEUs on the processor cache memory~\cite{shazli-TEST-2008}.
Meanwhile, continuous down-scaling of transistor feature size have increased the soft error rate per SRAM cell as well as the rate of MBUs~\cite{dixit2011impact,ibe2010impact,wilkening2014calculating}.
This challenge is getting more pronounced by the tremendous increase of cache memory size in the state-of-the-art processors, 
and the increased number of processor cores in the new 
generations of storage systems.

A large body of research investigates the effect of soft errors, including accelerated testing of SRAM and DRAM technologies~\cite{dixit2011impact,ibe2010impact,ogden2017impact}, field studies~\cite{dixit2009trends,sridharan2013feng}, and \emph{Architectural Vulnerability Factor} (AVF) and \emph{Mean Time To Failure} (MTTF) analysis~\cite{mukherjee2003systematic,sridharan2010using,wilkening2014calculating,suh2012macau}.
These studies shed light on the soft error problem from circuit level to micro-architecture and application level. 
At the micro-architecture/application level, these methods classify the outcomes of unmasked errors to two types of incidences, \emph{Detectable Unrecoverable Errors} (DUEs) and \emph{Silent Data Corruptions} (SDCs). 
This categorization, despite being useful, is insufficient for data storage systems, 
where the effect of soft errors should be classified to the failure types observable at the end-user side. 
Major types of storage failures affecting end-users are \emph{Data Loss} (DL)\footnote{DL is defined as the irreversible loss of stored user data.} and \emph{Data Unavailability} (DU)\footnote{DU is defined as the unavailability of storage system (hence, unavailability of user data) while the user data is not lost.}. 
However, 
the existing studies mostly focus on the soft error analysis in general purpose computing architectures that is not necessarily applicable to DSS, 
due to its unique hardware/software stack and dependability measurement requirements. 
This necessitates reviewing the applicability of existing art in the case of DSS. 

{In this paper, we offer the following contributions:}
\begin{itemize}
\item
{We propose a new metric, \emph{Storage System Vulnerability Factor} (SSVF), defined as the probability that a soft error results in DU/DL at the storage system level. This metric captures the dependability parameters of a storage system (DU and DL), as opposed to conventional AVF, which is primarily developed for traditional general-purpose computing architectures.}
\item
{Since the memory arrays are by far the most vulnerable components to soft errors~\cite{dixit2011impact,ibe2010impact}, using statistical fault injections we investigate the effect of soft errors in the cache memory of storage controller processors. For conducting our experiments, we implement the major functions of a typical storage controller, running on a full stack of storage system operating system, and use a full system simulator which is modified to simulate the hardware/software stack of DSS.}
\item
{The proposed analysis framework can cope with MBUs, technology dependent error characteristics such as bit error rate and the probability of MBU, different cache error protection schemes, and different redundancy architectures of storage controller.}
\item
{We carefully analyze the effect of soft errors at the cache memory, controller, and storage level to classify the failure cases (DU and DL) at the storage system level. Our analysis shows that 1) a considerable fraction of cache memory (by up to 40\%) holds the data of storage system users (called user data). Soft errors on such data can result in DL. 2) Soft errors in the rest of cache memory, occupied by Operating System (OS) and storage applications, at the worst case will result in DU, and in some rare cases may result in DL. 3) SDCs are the only causes of DL in the user side, while the contribution of SDCs in the storage unavailability is less than DUEs. This analysis concludes that conventional AVF is not a meaningful metric for demonstrating the susceptibility of data storage systems to soft errors.}
\item
{The effect of protection mechanism is investigated at both cache memory level and controller level. We examine different cache memory protections including linear Parity, interleaved Parity, linear \emph{Single Error Correction Double Error Detection} (SECDED), interleaved SECDED, and linear \emph{Double Error Correction Triple Error Detection} (DECTED). At the controller level, we examine single controller and dual controller configurations.}
\item
{The workload effect is studied by examining both synthetic and real workloads. The examined synthetic workloads capture the effect of different workload characteristics including randomness, inter-arrival time, and request size on DSS susceptibility to soft errors.}
\end{itemize}

The rest of this paper is organized as follows.
Section~\ref{sec:model} presents our proposed method for evaluating DU and DL using fault injections. 
Section~\ref{mathematical model} presents our proposed SSVF analysis and related discussions. 
Section~\ref{sec:result} presents our experimental results and observations.
Finally, Section~\ref{sec:conclusion} concludes the paper.

\section{Soft Error in DSS} 
\label{sec:model}

\subsection{Data Storage System Controller Simulation}
\label{sec:controller simulation}
The overall data flow in storage systems starts by receiving requests from end-user side which are queued by \emph{Front-End} logic through a storage network interface, using employed queue management algorithm.
\emph{Read} requests are responded by accessing the controller local cache, \emph{Global Memory} (GM), or disk subsystem, depending on the data residency.  
\emph{Write} requests, however, are typically responded after commitment on a mirrored GM, to assure the storage reliability.
A detailed description of data flow in DSS can be found in~\cite{gnanasundaram2012information,chang2008bigtable,nishtala2013scaling}.

\begin{figure}
    \centering
        \includegraphics[width=3.5in]{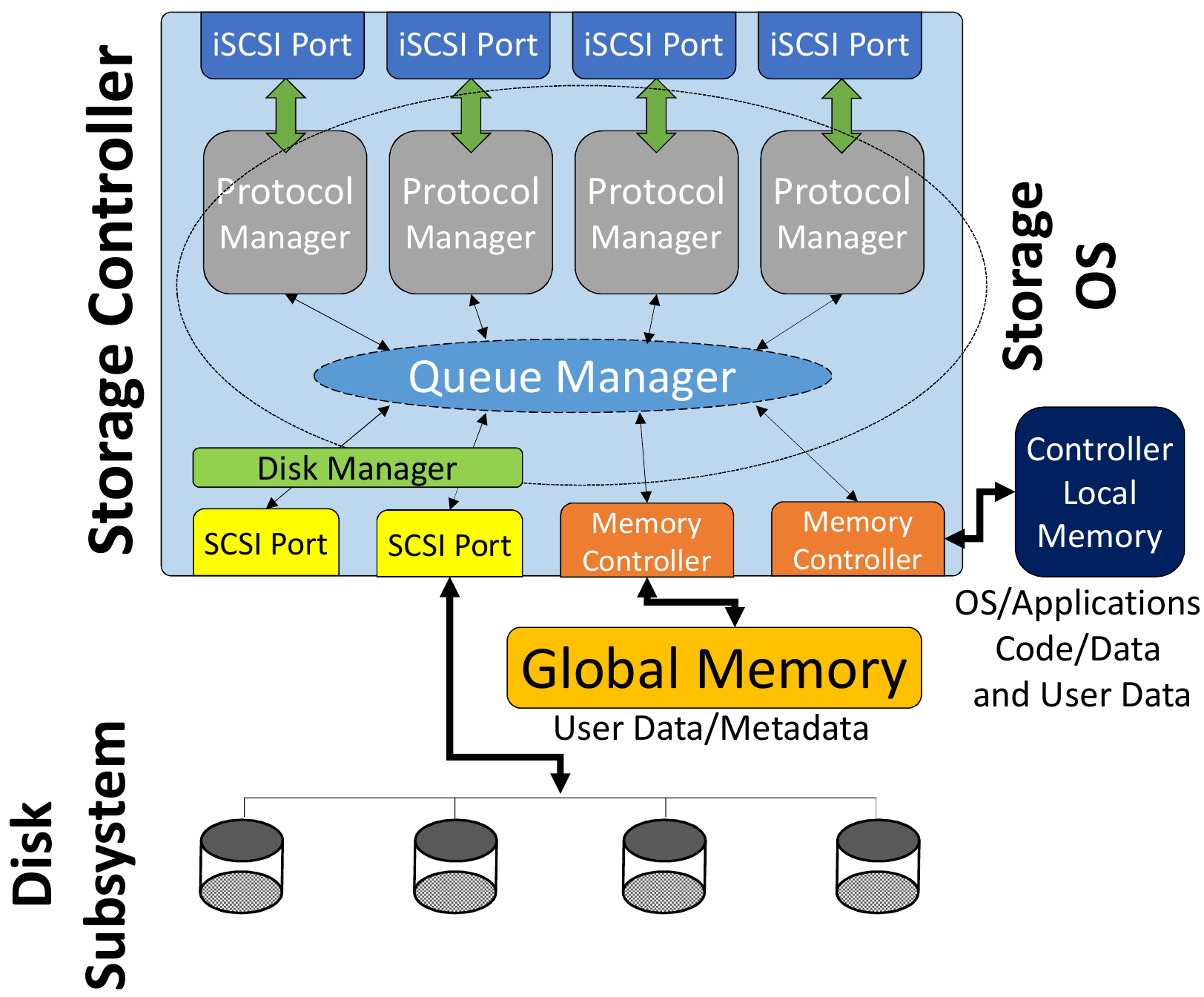}
    \caption{Hardware/software stack of simulated storage system.}
\label{fig:hardware-software-stack}
\end{figure}

We simulated the main task of a storage controller, including protocol management, request queue management, and disk management on the full 
stack of storage system operating system (Ubuntu 11.04, kernel version 2.6.38, ext4 file system) running on an X86 machine using MARSSx86 full system simulator~\cite{patel2011marss}. 
These tasks, servicing during mission time, are necessary to simulate the full stack of data flow in a real-world storage system. 
Figure~\ref{fig:hardware-software-stack} shows the overall hardware/software stack of the simulated storage system. 

MARSSx86 simulator enables us to simulate the full hardware stack, including processor, memory system, and I/O. 
MARSSx86 can also integrate with Disksim~\cite{bucy2008disksim} which simulates the behavior of disk subsystem and is able to simulate both disk and 
\emph{Solid Disk Drive} (SSD) arrays. 
For simulating GM, we define two memory controllers, one responsible for communication with local controller memory, that holds OS/applications code/data as well as user data input/output buffer. 
The second memory controller is communicating with GM which contains user data/meta-data. 
For disk subsystem we define one or multiple I/O ports, controlled by disk manager software. 
As our focus in this work is on the behavior of storage controller, and integrating MARSSx86 with Disksim increases the simulation time significantly, 
we simulate the effect of disk subsystem access by considering an average delay, statically obtained by Disksim.  
Finally, we define one or multiple iSCSI ports which are responsible for receiving the storage commands from the end-user and sending the responses.

A major component in storage software stack, as shown in Fig.~\ref{fig:hardware-software-stack}, is protocol management, which is responsible for receiving and responding the read/write requests via storage network protocols such as 
iSCSI and Fiber Channel.
Here we employ iSCSI protocol  
and use commercial IET~\cite{IET} tool to simulate the \emph{target} of the iSCSI protocol.
The iSCSI target, as named after, is the end-point of an iSCSI session that provides the input/output transfers initiated by iSCSI \emph{initiator}.
We initiate the requests, regarding the desired workload, via a remote \emph{Virtual Machine} (VM) through iSCSI protocol. 
These requests are received by our simulated storage controller machine. 
Each protocol manager thread is responsible for data communication of a single iSCSI port, while we can configure the controller with more than one port, each of which is handled by a separate thread.   
The second task is the request queue management, which is responsible for queuing the requests received from iSCSI port(s) and launching them to the 
disk manager or GM, regarding the storage system policies and data residency. 
The request queue management program has been developed in-house for \emph{First In First Out} (FIFO) algorithm.
Finally for disk management, we used Linux Generic SCSI~\cite{linuxgenericscsi}. 
Algorithm~\ref{pseudocode} shows the pseudo-code of data read/write stack in the DSS controller. 
This platform enables us to examine the DSS controller under desired workloads.

\begin{algorithm}
\caption{Storage Controller Pseudo Code}\label{pseudocode}
\begin{algorithmic}[1]
\fontsize{6}{6}\selectfont
\Procedure{Protocol Manager}{}
\State
\Call {Init(iSCSI)}{}
\While{$1$}
\State
\Call {Receive(Request)}{}
\State
\Call {Send(Response)}{}
\EndWhile
\EndProcedure
\Procedure{Queue Manager}{}
\State
\Call {Init()}{}
\While{$1$}
\State $currentRequest=$\Call {FIFO()}{}
\If {$currentRequest=Read$}
\If {$\textit{processor local cache hit}$}
\State $Data = \textit{Cache Access}$
\Else {\If {$\textit{GM hit}$} //GM: Global Memory 
\State $Data = \textit{GM Access}$
\Else {}{
\State $Data = \textit{Disk Access (SCSI)}$}
\EndIf}
\EndIf
\State iSCSI $ \gets \textit{Data} $
\State $\textit{Receive ACK}$
\Else {$\textit{ //currentRequest = Write}$}{
\State $Data \gets $ iSCSI
\State $\textit{GM Access //Write Data to GM}$
\State $\textit{Send ACK}$
}
\State \Call {Prefetch()}{}
\EndIf
\EndWhile
\Procedure{Disk Manager}{}
\State
\Call {Init(SCSI)}{}
\While{$1$}
\State
\Call {Receive(Disk Request)}{}
\State
\Call {Send(Response)}{}
\EndWhile
\EndProcedure
\State $\textbf{return } Statistics$
\EndProcedure
\end{algorithmic}
\end{algorithm}
\vspace{-0.3cm}

\subsection{Statistical Fault Injection}
\label{sec:statistical fault injection}

\begin{figure*}[ht]
    \centering
        \includegraphics[width=6.5in]{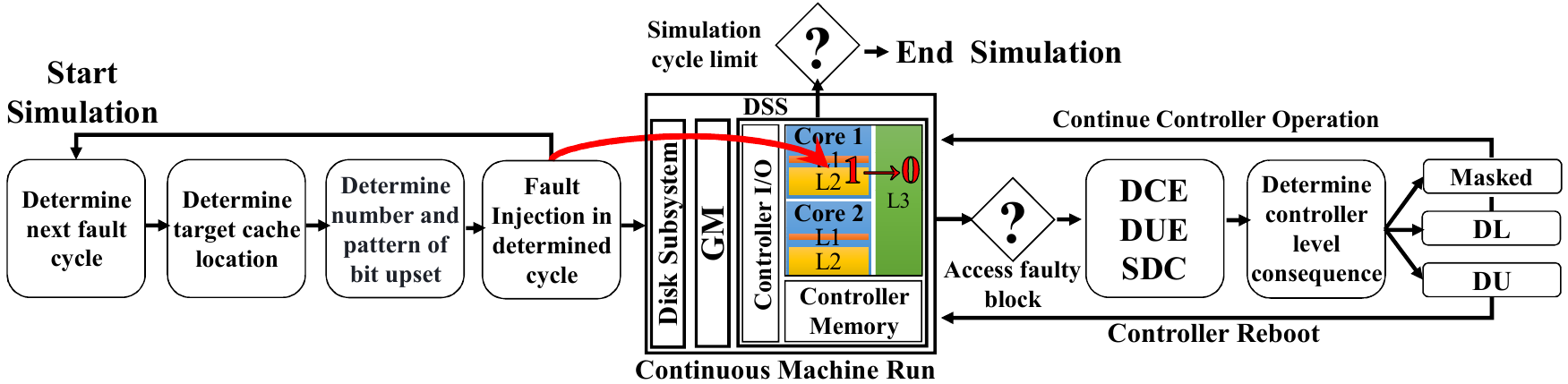}
    \caption{Fault Injection Scheme (Single Controller)}
\label{fig:fault-injection-scheme}
\end{figure*}

Fault injection on the simulated machine has been the subject of many works
~\cite{clark1995fault,wei2011comparing,ramachandran2008statistical,leveugle2009statistical,benso2003fault,arlat1993fault}.
These studies have developed the knowledge required to perform statistical fault injection in a simulated environment, which is adopted to our framework. 
Some fault injection extensions for known machine simulators are also developed, such as fault injection tool for Ruby~\cite{rubyfaultinjection}, 
and MaFIN for MARSSx86~\cite{7315134}.
However, MaFIN tool provides some trivial features and does not support our needs of detecting fault consequences from storage end-user sight. 
It also does not provide fault diagnostics in the resolution of individual threads, which is a necessity for our analysis. 
Moreover, this tool is not publicly available to download and is not supported by its developers any longer.

Fig.~\ref{fig:fault-injection-scheme} shows the overall flow of our fault injection procedure.
We modified the MARSSx86 simulator to add the possibility of fault injection. 
For each cache block, we add the fault information including the fault type (single bit-flip or MBU) and the location of erroneous bit(s).  
In our fault injection experiments we need to recognize whether the affected cache block belongs to the user or OS/application. 
Hence, we need to translate the physical memory address of affected block to the logical memory address, and then check the ownership of that logical address.  
To this end, we record the page table of simulated OS, and send it to the host OS (the OS that is hosting MARSSx86), 
using a facility of MARSSx86, named PTLCall.
As the page table is dynamically changing during runtime, we need to send the page table back 
at the same machine cycle the fault is injected and recognize the ownership of the faulty block.

As shown in Fig.~\ref{fig:fault-injection-scheme}, the machine cycle of fault injection, as well as the address and spatial characteristics of the injected fault, are determined 
depending on the desired rate of soft error and the probability of MBU. 
We describe the target error bit patterns in Section~\ref{sec:error bit pattern}.
On every access to a cache word, we check the status of tag and data and take further actions if it is faulty. 
Depending on the protection scheme, error bit pattern, and the cache access type (read or write), the ECC may correct, detect, or not detect the error. 
For each error case, the controller takes suitable actions clarified in Section~\ref{sec:processor-level}. 
We finally record the failure (DU/DL) statistics.

\subsection{Cache Architecture and Error Bit Pattern Assumptions}
\label{sec:error bit pattern}
{Hereby we note the assumption we took about processor cache protection scheme and the error bit pattern. 
Note that in the models described in this work and all the simulations, we assume MESI\footnote{MESI is an acronym representing four possible cache line  states, 
Modified (M), Exclusive (E), Shared (S), and Invalid (I).} write-back cache replacement policy~\cite{papamarcos1984low} for both L1 and L2 caches. }

{The effect of MBU in the cache memory depends on the employed error protection scheme and the error bit pattern (also called as fault geometry)~\cite{wilkening2014calculating}. 
Error protection schemes such as parity and ECC can be linear or interleaved (interleaving can be logical, way physical, and index physical~\cite{wilkening2014calculating}). 
In this work we consider linear ECC, which protects a data word via a single ECC word along all bits, and logical \emph{k-way} interleaved ECC, which splits data word into $k$ interleaved ECCs. 
The study by~\cite{wilkening2014calculating} shows that using logical interleaving results in many times lower vulnerability factor than that of physical 
interleaving. }

\begin{figure}
    \centering
        \includegraphics[width=3.5in]{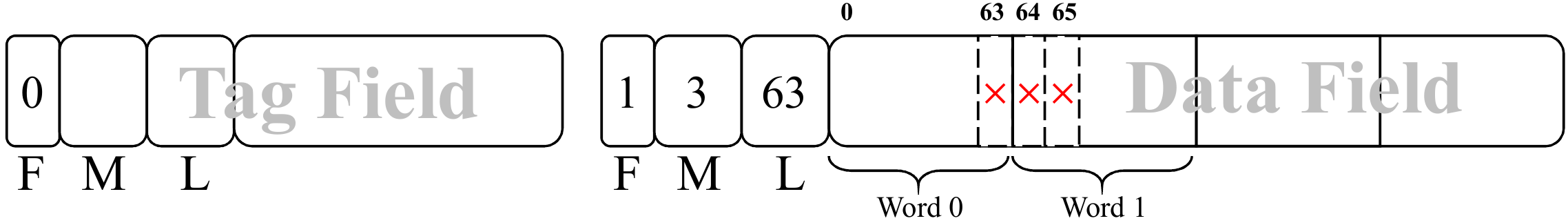}
    \caption{{Recording fault attributes for tag and data fields of each cache block. Each cache block has a 32-bit tag field and 512 bit data field, while data field is divided to 8 words of 64-bit. The $F$ field is $1$ when the block tag/data is faulty. The $M$ field shows the size of contiguous error and $L$ is the location of the first faulty bit. For example, in this figure the tag $F$ field is $0$, indicating that the tag field is fault free. Meanwhile, data $F$ field is $1$, indicating that the data field is faulty. $M=3$ shows that the size of contiguous error is $3$ and $L=63$ shows the location of the first faulty bit. Hence, the bits $63$, $64$, and $65$ are faulty. Assuming that each data word size is 8 bytes, this fault targets two adjacent data words, \emph{Word 0} and \emph{Word 1}, known as MCU~\cite{de2016evaluation}. As we assume that each data word is protected with a separate ECC, this error bit pattern is interpreted to a single error in \emph{Word 0} and a double error in \emph{Word 1}.}}
\label{fig:fault metadata}
\end{figure}

{Error bit pattern can be contiguous and non-contiguous with different size and geometries~\cite{ibe2010impact}. 
The focus of this work is on the most common and problematic pattern, contiguous $M\times1$ pattern~\cite{ibe2010impact,wilkening2014calculating}, which modifies 
$M$ contiguous bits in a word line. 
Hence, the spatial characteristic of each fault incidence in a cache line can be recorded by the location of the first faulty bit $L$ and the size of contiguous error $M$. 
We assume the probability that a cache block is affected more than one time in a fault injection experiment is zero (note that it also never happened in our fault injection experiments). 
Hence, we can ignore the possibility of the fault accumulation and record one fault incidence per cache data field and cache tag field. Fig~\ref{fig:fault metadata} shows how we record the fault attributes for each cache block.}

\subsection{Impact of Soft Errors at Controller Level}
\label{sec:processor-level}
\begin{figure}
    \centering
        \includegraphics[width=3in]{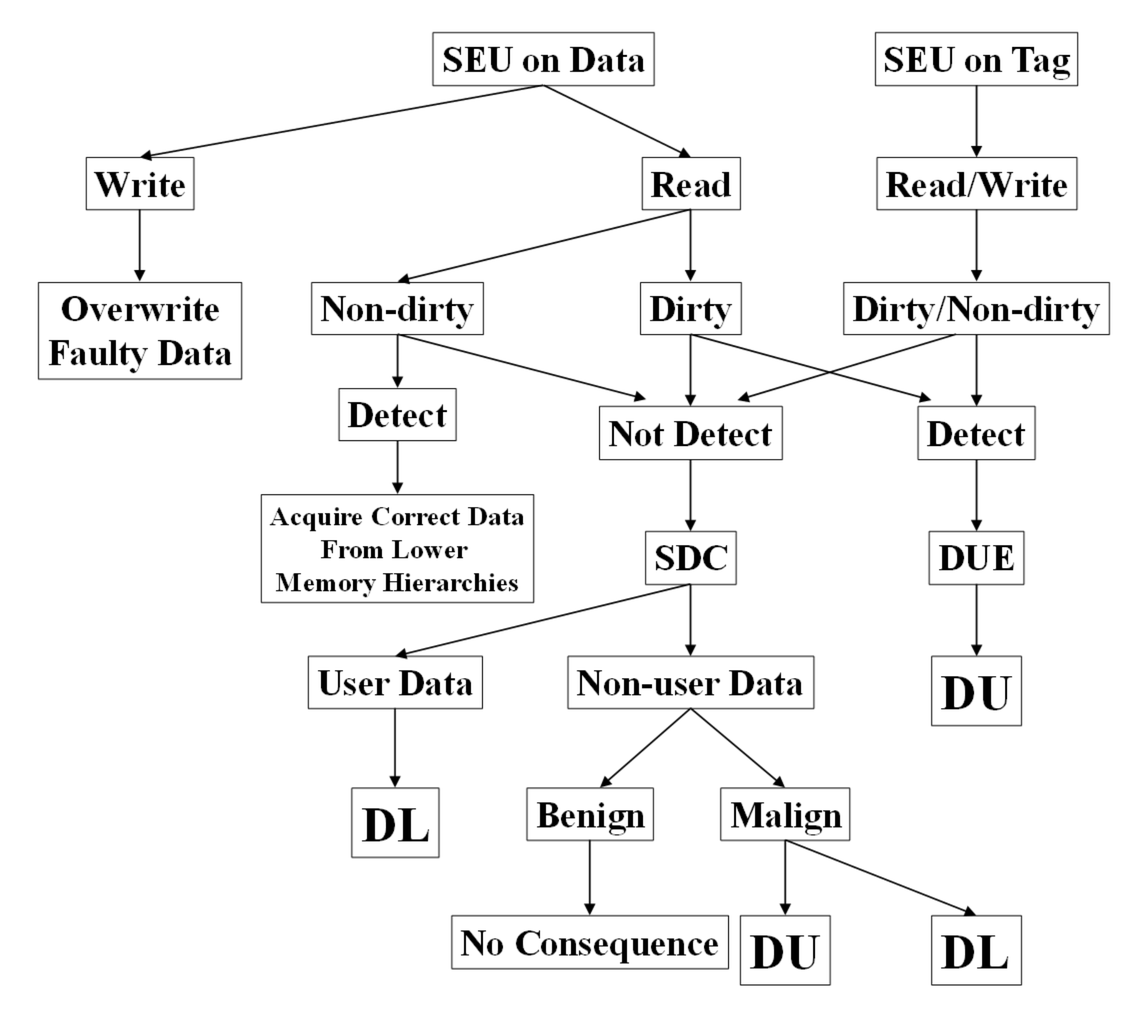}
       
    \caption{{Impact of soft errors at Controller Level (Single Controller)}}
  \label{fig:fault-injection-controller-level}
\end{figure}

The soft error in a cache block can propagate to the controller, and further manifest itself at the storage level as DU and DL, 
under different cache access and error characteristic scenarios, which is analyzed next.  
Fig.~\ref{fig:fault-injection-controller-level} summarizes the different error cases and the corresponding 
outcomes at the controller level upon an access to a faulty cache block.
As the consequences of errors in tag field and data field are different, we separate them in our analysis.
The errors that are correctable, detectable, and not detectable by ECC are called \emph{Detectable Correctable Error} (DCE), DUE, and SDC, respectively.
Note that in the case of DCE, the error can be corrected immediately and there will be no further consequences. 

\subsubsection{Read Access to Non-dirty Blocks}
\label{sec:read-non-dirty}

Upon a read access to a non-dirty cache block with faulty \emph{data} field,    
if the error is detected (DUE), the correct data can be fetched from lower memory hierarchies. 
Otherwise, in case of SDC (undetected error), the error results in DL, if 
happened on a user block.
{In the case that SDC happens on a non-user block (belonging to OS/applications data), there is a possibility that the SDC is either \emph{Not Activated} (if the SDC targets a code location never accessed by OS/applications) or \emph{Not Manifested} (if the SDC is activated but does not manifest as system level failure, with $P_{NotManifest}$), called \emph{Benign} SDC. However, the SDC can possibly result in software malfunctions, finally resulting in controller reboot. In some rare conditions, the OS/application malfunctions can even harm the end-user data\footnote{{Gu et al. performed a deep characterization of Linux kernel behavior under errors~\cite{gu2003characterization}.}}, resulting in DL (with $P_{OS_{DL}}$).}

{Upon a read access to a non-dirty cache block with faulty \emph{tag} field, the controller cannot assure whether the faulty block was originally dirty or not. In {that} case, the controller takes the most conservative action and reboots itself, no matter the error happened on the user data or non-user data. Note that the occurrence of DUE on the user data does not result in DL if the controller is immediately rebooted after the cache access. {The reason behind is} the conventional behavior of DSS controllers in responding to data write requests. For the sake of reliability, the write requests are responded once the data is written back to the \emph{Global Memory} (GM) of the storage system. Hence, if the controller is rebooted in the case of DUE, the write request is not responded and the user needs to resubmit its request, {preventing DL}.}

\subsubsection{Read Access to Dirty Blocks}
\label{sec:read-dirty}
{Upon a DUE on the data/tag field of a dirty cache block (Fig.~\ref{fig:fault-injection-controller-level}), the {correct data is not obtainable} from lower memory hierarchies. {In that case}, the controller reboots itself to remove the chance of data loss, no matter the error happened on the user data or non-user data. Finally, in the case of SDC {on dirty blocks the controller takes the same action as non-dirty blocks.}}

\subsubsection{Write Access}
Upon a write access to a cache word with faulty \emph{data} field, the faulty word would be replaced by the new word 
and the error is masked.
In the case of write access to a cache block with faulty \emph{tag} field, however, the following actions are taken; 
Suppose that address $A$ (containing $D_A$) is changed to address $B$ (containing $D_B$) due to an error on the tag field. 
After the error occurrence, the address $B$ contains $D_A$, which is wrong.  
Write access to address $B$ replaces just one word of the wrong data ($D_A$) with the new correct data word ($D_B'$), while the rest of address $B$ still contains 
the wrong data $D_A$. 
Moreover, the block $A$ is now disappeared from the cache memory;
As $A$ does not exist in the cache memory (as its address have changed to $B$), the controller has no information 
whether $A$ was originally dirty or non-dirty. 
Hence, if the error in the tag field is detected (DUE), for the sake of storage reliability, the controller takes the conservative assumption that $A$ was originally dirty.
Consequently, similar to the case of DUE on read accesses, a system reboot is necessary to prevent DL (as discussed in Section~\ref{sec:read-dirty}). 
Finally in the case of SDC (undetected error) on the tag field, the consequences and further actions depend on whether the affected block belongs 
to user data or OS/applications, as discussed in Section~\ref{sec:read-non-dirty}.

\subsection{Impact of Soft Errors at the System Level} 
\label{sec:system-level}
In our study, we investigate two common architectures, single controller and 
\emph{Dual Initiated} (DuIn) controller for the storage system. 
However, other redundancy architectures such as \emph{Triple Modular Redundancy} (TMR) and \emph{N-Modular Redundancy} (NMR) can also be analyzed similarly. 

\subsubsection{Single Controller}
In the case of single controller, during the time the controller 
is being rebooted (discussed in Section~\ref{sec:processor-level}), 
the storage system is unavailable (DU). 
We aggregate all DU periods during system simulation. 
The unavailability of the system would be the fraction of this value 
over the total simulation time.
The DL per simulation is also reported by the aggregation 
of all DL events (number of lost bytes) happened during the simulation.
In calculating DL bytes, one has to consider the difference between the consequences of SDC on tag and data fields, separately. 
{{SDC on} the data field results {in DL} of one data word, or eventually two adjacent data words, known as \emph{Multiple Cell Upset} (MCU)~\cite{de2016evaluation}, {while} SDC on the tag field results in DL of the entire block.}

\subsubsection{Dual Controller}
\label{sec:systemlevel-dual}
In the case of dual initiated controllers, two controllers are  
simultaneously and independently running.
In the case one controller is being rebooted, the other operating 
controller takes over the tasks. 
However, {when the failure of dual initiated controllers coincide,} the storage system is unavailable (DU).   
Similar to the case of single controller, {total DU and} DL per simulation {is} obtained by the aggregation of {individual} DU and DL incidences, respectively.

\section{Storage System Vulnerability Factor (SSVF)}
\label{mathematical model}
Using our storage simulation and fault injection method, in this section we analyze AVF of storage controller and 
show that AVF analysis is not sufficient to quantify the DU and DL.
We further propose a new metric, SSVF, to better represent the effect of soft errors in  
storage systems.
\subsection{AVF Analysis}
{AVF is defined as the fraction of faults (soft errors in our case) that become errors.
We define $AVF^{SDC}$ as the fraction of faults that {leads to} SDC, and $AVF^{DUE}$ as the fraction of faults that {leads to} DUE.}
As discussed in Section~\ref{sec:model}, SDC has different consequences, regarding the fault location (either on tag or data field) and whether it belongs to {user or OS/application (non-user)}.
Hereby, we propose differentiating $AVF^{SDC}$ {regarding the fault location (tag and data field) and data ownership (user and non-user).}
We define $AVF^{SDC}$ of faults happening on \emph{Tag Field} (TF) of \emph{User Data} (UD) ($AVF^{SDC}_{TF_{UD}}$) as follows:
\begin{equation}
\label{equ:avf-sdc-user-tag}
\begin{split}
\resizebox{1\hsize}{!}{$AVF^{SDC}_{TF_{UD}}=\frac{\sum\limits_{i=1}^N [SDCs~ on~ TF~ of~ UD~at~cycle~i] ~(SDC_{TF_{UD}})}{\sum\limits_{i=1}^N [ Faults~ injected ~on~ TF~ of~ UD ~ at ~ cycle ~ i]}$} 
\end{split}
\end{equation}

Where $N$ is the number of machine cycles of storage {service}. 
{Similarly, we define $AVF^{SDC}_{DF_{UD}}$ as the fraction of faults in \emph{Data Field} (DF) of user data that {leads to} SDC.}
{$AVF^{SDC}_{TF_{NUD}}$ is also defined as the fraction of faults in tag field of \emph{Non-User Data} (NUD) that {leads to} SDC.} 
{{Similarly,} $AVF^{SDC}_{DF_{NUD}}$ is defined as the fraction of faults in data field of non-user data that {leads to} SDC.}

The analysis of storage system failure breakdown in Section~\ref{sec:model} shows that {DUE} results in {controller} reboot, no matter  
the fault occurs on the user data or non-user data. 
{Hereby we define $AVF^{DUE}_{TF}$ as the fraction of faults on tag field that {leads to} DUE, and $AVF^{DUE}_{DF}$ as the fraction of faults on data field that {leads to} DUE}, as follows: 
\begin{equation}
\label{equ:avf-due-tag}
\begin{split}
\resizebox{1\hsize}{!}{$ AVF^{DUE}_{TF} = \frac{\sum\limits_{i=1}^N [DUEs~ on~ TF~ at ~ cycle ~i] ~(DUE_{TF})}{\sum\limits_{i=1}^N [Faults~ injected ~on~ TF~at~cycle~i]}$}
\end{split}
\end{equation}

In Section~\ref{sec:results-avf-analysis}, we present AVF values obtained by fault injection experiments for different cache protection schemes, and 
show that it cannot represent the DU/DL of the storage system, as DU in storage system is caused by both SDC and DUE events at the controller level. In the next section, we present SSVF that projects the effect of soft errors on DU/DL at the storage system level, rather than DUE and SDC at the controller level. 

\subsection{SSVF Analysis}
While $AVF^{SDC}$ and $AVF^{DUE}$ are defined as the processor vulnerability to SDC and DUE, 
in the case of storage systems, the storage vulnerability can be defined as the fraction of soft errors resulting in DU and DL. 
Hereby, we define $SSVF^{DL}$ as the probability that soft error in cache memory results in DL at the storage level.
Similarly, we define $SSVF^{DU}$ as the probability that soft error in cache memory results in DU at the storage level.
Modeling $SSVF$ in terms of $AVF^{DUE}$ and $AVF^{SDC}$ is very challenging, 
as the analysis in Section~\ref{sec:model} as well as the results of Section~\ref{sec:results-avf-analysis} show that DU and DL are caused by both DUEs and SDCs, and AVF cannot address the final consequence of a soft error at the storage level.

{As our analysis in Section~\ref{sec:model} shows (also confirmed by the results provided in Fig.~\ref{fig:protections}), errors on tag and data fields have different system level consequences and different chance to result in a failure.}
Hence, we define different SSVF values for tag and data fields.

We define $SSVF^{DL}_{TF}$ as the probability that a soft error in the tag field of cache memory results in DL according to Equation~\ref{equ:svf-dl-tag}.
\begin{equation}
\label{equ:svf-dl-tag}
\begin{split}
\resizebox{0.75\hsize}{!}{$SSVF^{DL}_{TF}=\frac{\sum\limits_{i=1}^N  [DL~on~ TF~ at~cycle~i]}{\sum\limits_{i=1}^N  [Faults~ injected ~on~ TF~ at~cycle~i]}$} 
\end{split}
\end{equation}

Similarly, we define $SSVF^{DL}_{DF}$ as the probability that a soft error in the data field of cache memory results in DL.
We also define $SSVF^{DU}_{TF}$ as the probability that a soft error in the tag field of cache memory results in DU according to Equation~\ref{equ:svf-du-tag}.
Similarly, we define $SSVF^{DU}_{DF}$ as the probability that a soft error in the data field of cache memory results in DU.
\begin{equation}
\label{equ:svf-du-tag}
\begin{split}
\resizebox{0.75\hsize}{!}{$SSVF^{DU}_{TF}=\frac{\sum\limits_{i=1}^N  [DU~on~ TF~ at~cycle~i]}{\sum\limits_{i=1}^N  [Faults~ injected ~on~ TF~ at~cycle~i]}$} 
\end{split}
\end{equation}

\section{Results and Observations}
\label{sec:result}
\subsection{Experimental Setup}
\label{experimentalsetup}
In the experiments presented in this section, we assume each controller has one processor with four \emph{Out-of-Order} (OoO) cores with shared L2 cache memory with 45nm technology node.
The system configuration, as well as the configuration of each core, cache memory, main memory, and interconnections is appeared in Table~\ref{tab:configuration}. 
For the rate of MBU, we used the MBU rate reported by Dixit et al.~\cite{dixit2011impact} for 45nm technology node (1-bit: 62\%, 2-bit: 25\%, 3-bit: 7\%, 4-bit: 6\%).
The cache protections include linear parity (parity), linear SECDED, two-way interleaved parity, 
two-way interleaved SECDED, and linear DECTED. 
{We obtain the suitable number of fault injection experiments per processor/workload configuration, $n$, using the approach presented by Leveugle et al.~\cite{leveugle2009statistical}.
As Leveugle et al. suggest, the number of fault injections ($n$) is computed {using} Equation~\ref{equ:number-of-injected-faults}~\cite{leveugle2009statistical}.   }
\begin{equation}
\label{equ:number-of-injected-faults}
\begin{split}
n=\frac{N}{1+e^{2} \times \frac{N-1}{t^{2}\times p \times (1-p)}}
\end{split}
\end{equation}

{Where $N$ is the population of faults (all possible fault incidences in different time/location, infinite in our case), $p$ is the estimated probability of faults resulting a failure (as this value is usually unknown, it is recommended to take the most conservative value, $p=0.5$, as we did), $e$ is the margin of error, and $t$ is the cut-off point corresponding to the confidence {level} with respect to {\emph{Normal}} distribution.  
Assuming $N=\infty$, we evaluate $n$ by considering 1\% error interval ($e=0.01$), confidence level of 95\% ($t=1.96$), and the most conservative value for $p$ ($p=0.5$) that results in $n=9604$. Accordingly, we conduct $10,000$ fault injection experiments per configuration. 
The {fault injection is} verified to have uniform distribution {over} both time and location.}

\subsection{{Fault Tracking}}
\label{sec:fault-tracking}
{{In the proposed fault injection environment, we} need to carefully track the sequence of accesses to the {faulty cache blocks} to {accurately extract} processor-level and system-level {failure} statistics. The injected faults are tracked in the following steps:}
\begin{itemize}
\item
{\textbf{Fault Injection:} Once the fault is injected, statistics such as target cache hierarchy/number, target cache line (way and set), whether fault is on tag or data, type of MBU (number of bit flips), fault location in the cache line, and whether the fault targets user data, OS/application data, or an invalid cache line, is recorded. }
\item
{\textbf{Initial DL Collection:} Once the fault results in SDC and targets user data, an initial DL is possible {in} some scenarios.}
\item
{\textbf{DL Propagation:} The DL propagation is recorded once the faulty data is {either} read or written back to the lower memory {hierarchy}.}
\item
{\textbf{Fault Masking:} We carefully consider the fault masking scenarios upon cache write, cache update, cache evict, processor reboot, ECC correction, and ECC detection when {the} cache line is clean. }
\item
{\textbf{SDC, DUE, and DCE Incidences:} Upon an access to a faulty cache line, the incidence of SDC, DUE, and DCE is recorded\footnote{{Please note that we follow the definition of SDC suggested by Mukherjee et al.~\cite{mukherjee2005soft}, in which both SDC and DUE occur as an outcome of faulty cache access. Hence, a never accessed faulty cache line is not counted in our SDC/DUE stats.}}.}
\end{itemize}

\subsection{Examined Workloads}
\label{sec:examinedworkloads}
To measure the effect of different storage workloads, we conduct our fault injection experiments for synthesized and real workloads. 
The synthetic workloads are attributed by \emph{Inter Arrival Time}, defined as the average time between two successive requests, \emph{Request Size}, defined as the average size of the requests, and \emph{Randomness}\footnote{A recent study presented in~\cite{tarihi2016hybrid}  classifies the storage I/O to \emph{Sequential}, \emph{Overlapped}, and \emph{Strided}. A sequence of requests is recognized as \emph{Random} when it does not follow sequential, overlapped, and strided characteristics.}. 
The synthetic workload includes different inter-arrival times (average of 10, 100, and 1000 microseconds with exponential distribution), different request size (average of 1, 10, 100, and 1000 kilobytes with exponential distribution), and different randomness (sequential requests versus random requests, while the random request address is generated with uniform distribution over all storage space). 
{The real workloads include \emph{Financial\_1} and \emph{Financial\_2} (I/O traces from OLTP applications running at two large financial institutions)~\cite{umasstrace} and \emph{Websearch\_1}, \emph{Websearch\_2}, and \emph{Websearch\_3} (I/O traces from a popular search engine)~\cite{umasstrace}.}
The software stack which handles the storage workloads is described in Section~\ref{sec:controller simulation}.

\begin{table}
\centering
\caption{MARSSx86 Simulation Configuration}
\begin{center}
\begin{adjustbox}{width=\textwidth/2,totalheight=\textheight,keepaspectratio}
    \begin{tabular}{ | c | c | c | c |}
    \hline
\cellcolor[HTML]{656565}{\color[HTML]{FFFFFF} \textbf{machine}}	&    st\_FUs: 2	&    pending\_queue\_size: 256	&    type: dram\_cont \\ \hline
  name: shared\_l2	&    frontend\_width: 4	&    coherence: MESI	&    RAM\_size: 134217728 \\ \hline
  cpu\_contexts: 4	&    dispatch\_width: 4	& \cellcolor[HTML]{656565}{\color[HTML]{FFFFFF} \textbf{L1\_D\_0}}	&    number\_of\_banks: 64 \\ \hline
  freq: 1600000000	&    issue\_width: 4	&    type: cache	&    latency: 80 \\ \hline
 \cellcolor[HTML]{656565}{\color[HTML]{FFFFFF} \textbf{ooo\_0\_0}}	&    writeback\_width: 4	&    size: 131072	&    latency\_ns: 50 \\ \hline
  type: core	&    commit\_width: 4	&    sets: 256	&    pending\_queue\_size: 128 \\ \hline
    threads: 1	&    max\_branch\_in\_flight: 24	&    ways: 8	&  \cellcolor[HTML]{656565}{\color[HTML]{FFFFFF} \textbf{p2p\_core\_L1\_I\_0}} \\ \hline
    iq\_size: 64	&    per\_thread:	&    line\_size: 64	&    type: interconnect \\ \hline
    phys\_reg\_files: 4	&      rob\_size: 128	&    latency: 2	&    latency: 0 \\ \hline
    phys\_reg\_file\_int\_size: 256	&      lsq\_size: 96	&    pending\_queue\_size: 256	&  \cellcolor[HTML]{656565}{\color[HTML]{FFFFFF} \textbf{p2p\_core\_L1\_D\_0}} \\ \hline
    phys\_reg\_file\_fp\_size: 256	&\cellcolor[HTML]{656565}{\color[HTML]{FFFFFF} \textbf{core\_0\_cont}}	&    coherence: MESI	&    type: interconnect \\ \hline
    phys\_reg\_file\_st\_size: 48	&    type: core\_controller	& \cellcolor[HTML]{656565}{\color[HTML]{FFFFFF} \textbf{L2\_0}}	&    latency: 0 \\ \hline
    phys\_reg\_file\_br\_size: 24	&    pending\_queue\_size: 128	&    type: cache	& \cellcolor[HTML]{656565}{\color[HTML]{FFFFFF} \textbf{p2p\_L2\_0\_MEM\_00}} \\ \hline
    fetch\_q\_size: 48	&    icache\_buffer\_size: 32	&    size: 2097152	&    type: interconnect \\ \hline
    frontend\_stages: 4	&\cellcolor[HTML]{656565}{\color[HTML]{FFFFFF} \textbf{L1\_I\_0}}	&    sets: 4096	&    latency: 0 \\ \hline
    itlb\_size: 32	&    type: cache	&    ways: 8	& \cellcolor[HTML]{656565}{\color[HTML]{FFFFFF} \textbf{split\_bus\_00}} \\ \hline
    dtlb\_size: 32	&    size: 131072	&    line\_size: 64	&    type: interconnect \\ \hline
    total\_FUs: 8	&    sets: 256	&    latency: 5	&    latency: 6 \\ \hline
    int\_FUs: 2	&    ways: 8	&    pending\_queue\_size: 128	&    arbitrate\_latency: 1 \\ \hline
    fp\_FUs: 2	&    line\_size: 64	&    config: writeback	&    per\_cont\_queue\_size: 16 \\ \hline
    ld\_FUs: 2	&    latency: 2	& \cellcolor[HTML]{656565}{\color[HTML]{FFFFFF} \textbf{MEM\_0}} & 	 \\ \hline

    \end{tabular}
    \end{adjustbox}
\end{center}

\label{tab:configuration}
\end{table}

\subsection{AVF Analysis}
\label{sec:results-avf-analysis}
\begin{figure}
\begin{centering}
\includegraphics[width=0.5\textwidth]{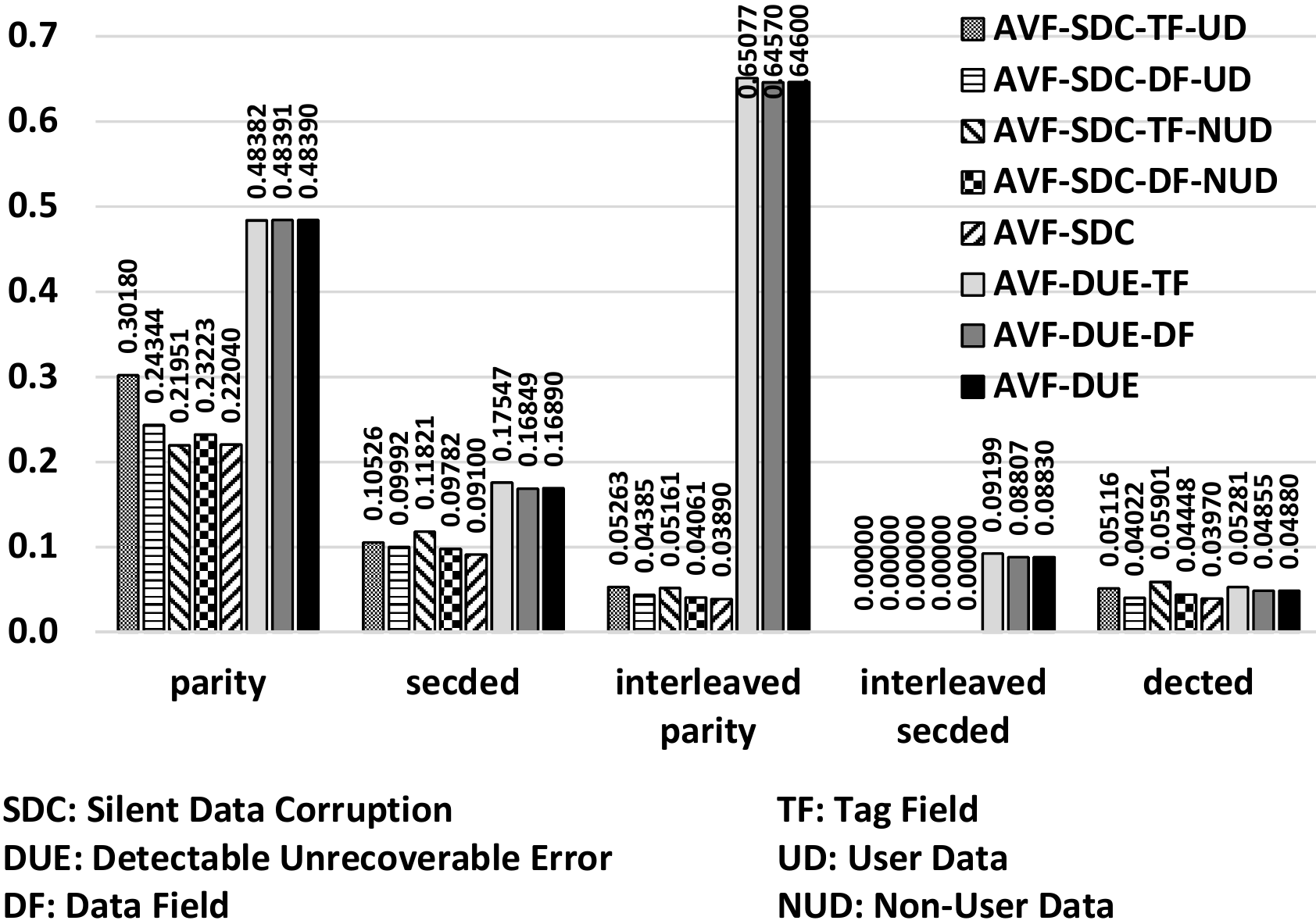}
\caption{AVF for Different Cache Error Protection Schemes (Financial\_1 workload)}
\label{fig:protections}
\par\end{centering}
\end{figure}

Fig.~\ref{fig:protections} {shows AVF values evaluated by using fault injection experiments} for different cache protection schemes (under Financial\_1 workload). 
{As shown in Fig.~\ref{fig:protections}, different protection schemes result in totally different AVF values, regarding their differences in detection/correction capability. 
{The results show that the tag fields targeted by DUEs have almost the same vulnerability as data fields. In the case of SDCs, however, tag fields show a slightly greater vulnerability than data fields.}
{This observation is described by the fact that SDCs on the data field and tag field have different sources of masking, while the masking in the data field is more effective. SDCs on the tag field may possibly have no consequence if the affected cache line is originally non-dirty. In that case, if the soft-error changes the tag value to an invalid memory address or an address that has never been accessed in runtime, the fault is masked. Meanwhile, SDCs on the data field just pollute one single word (or two adjacent words), while the SDCs on the tag field pollute the entire cache line. In the former case, the possibility of fault masking is more, as there is a chance that the faulty word is either never accessed or overwritten.}
Moreover, Fig.~\ref{fig:protections} shows zero $AVF^{SDC}$ values for interleaved SECDED. 
The reason is that in our fault injection {experiments}, the largest MBU is 4-bit upset. 
Hence, the two way interleaved SECDED can {either} correct or detect all the errors and there is no chance {of} SDC.  
{Another observation from AVF results (Fig.~\ref{fig:protections}) is that AVF does not represent the chance of DU and DL in data storage systems. As an example, $AVF^{SDC}$ for parity code is 0.22, meaning that 22\% of soft errors lead to SDC when using parity code. Meanwhile, our experiments show that only 2\% of soft-errors result in DL. Hence, the SDC reported by AVF metric is one order of magnitude greater than the actual DL at the end-user side.}
}

\subsection{SSVF Analysis}
\label{sec:SSVF Analysis}
\begin{figure*}
    \centering

        \includegraphics[width=0.7\textwidth]{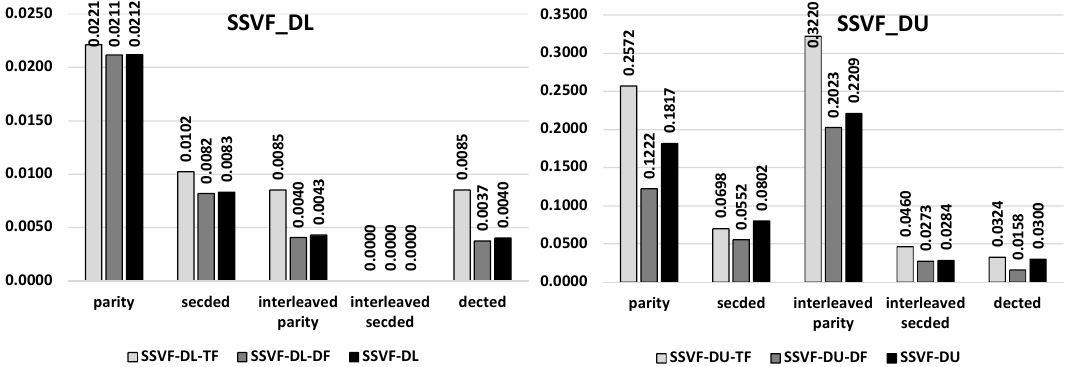}

    \caption{$SSVF^{DU}$ and $SSVF^{DL}$ for different cache protection schemes (Financial\_1 workload)}
    \label{fig:svfdudl}
\end{figure*}

{Fig.~\ref{fig:svfdudl} shows the $SSVF$ values for different cache protection schemes (under Financial\_1 workload). By definition, $SSVF^{DU}$ and $SSVF^{DL}$ values are representing the fraction of SEU events resulting in DU and DL, respectively. The results show that parity protection has the highest chance of DL, due to having the greatest $SSVF^{DL}$ value. Moreover, for all protection schemes the $SSVF^{DL}_{TF}$ is slightly greater than $SSVF^{DL}_{DF}$, showing that SEUs on the tag field have more chance to result in DL, {compared to SDCs on the data field}. We can describe this observation by the way SDC is propagated in tag and data field. In the case SDC targets the cache data field, there is a possibility it is shared between two {adjacent} data words and {goes} detected/corrected in each individual word. Hence, in those cases the SDC may have less impact on data field compared to tag field. 

{After} parity, SECDED has the {second} greatest $SSVF^{DL}$ value, followed by interleaved parity and DECTED {that show very near $SSVF^{DL}$}. Interleaved parity and DECTED both have equal detection capabilities (both cannot detect 4-bit MBUs). {Hence,} it was expected that both {protections} perform similar in terms of DL, while the experiment results also confirmed our expectation (interleaved parity and DECTED respectively had 43 and 40 initial DL incidences). Finally, we observed zero $SSVF^{DL}$ for interleaved SECDED, as interleaved SECDED can detect up to 4-bit MBUs, hence, no chance of DL.}

$SSVF^{DU}$ results, however, do not have the same trend as $SSVF^{DL}$. The greatest $SSVF^{DU}$ value belongs to interleaved parity. 
{Interleaved parity has no correction capability, but when considering $M\times 1$ error bit pattern (as described in Section~\ref{sec:error bit pattern}) it has a higher detection capability than both parity and SECDED.}
Hence, interleaved parity is expected to have {a} lower $SSVF^{DL}$ {than both parity and SECDED}, as the chart shows. Meanwhile, due to having no correction capability, all detected faults (DUEs) result in DU. In specific, 2-bit MBUs on user data lead to DL in parity protection, while in {the} case of interleaved parity they are detected and result in DU. Similarly, 3-bit MBUs on user data lead to DL in SECDED protection, while interleaved parity can detect 3-bit MBUs, resulting in DU. Interleaved SECDED and DECTED protections both perform better than interleaved parity in terms of error correction, while interleaved SECDED has also {a} better detection capability {than both DECTED and interleaved parity} ({interleaved SECDED} can detect up to 4-bit MBUs). So it was expected that both interleaved SECDED and DECTED have better $SSVF^{DU}$ and $SSVF^{DL}$ than interleaved parity, as the results show. 


\subsection{Comparison of Cache Protection Mechanisms}

\begin{figure}
    \centering
 
	\subfigure[Financial\_1 Workload] 
	{
        \includegraphics[width=0.5\textwidth]{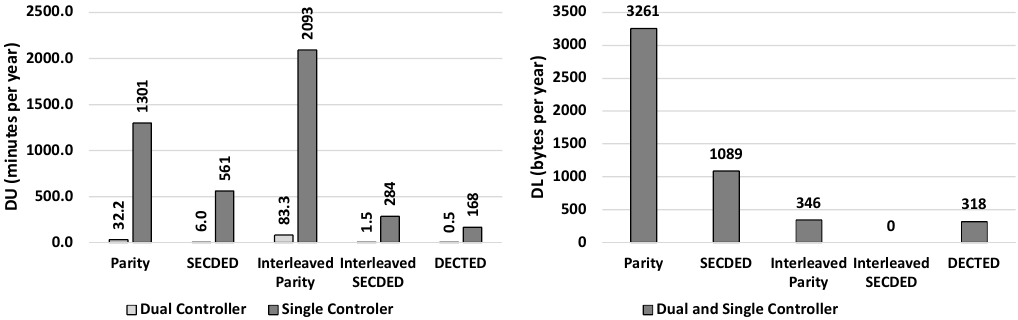}
        \label{fig:dudl-financial1}
    }
	\subfigure[Financial\_2 Workload] 
	{
        \includegraphics[width=0.5\textwidth]{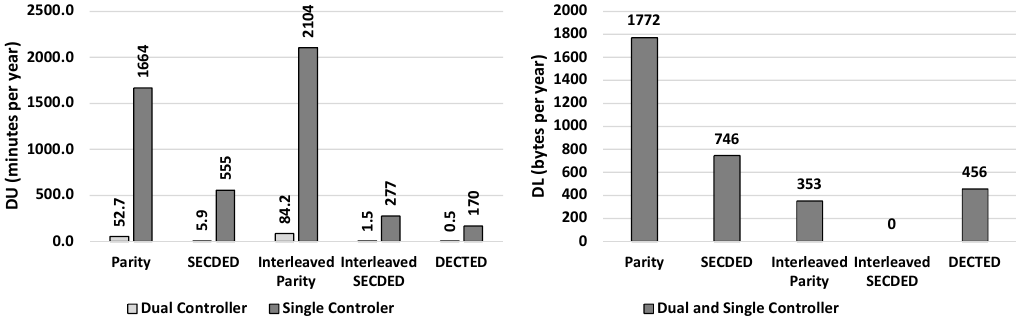}
        \label{fig:dudl-financial2}
    }    
    \subfigure[Websearch\_1 Workload]
	{
		\includegraphics[width=0.5\textwidth]{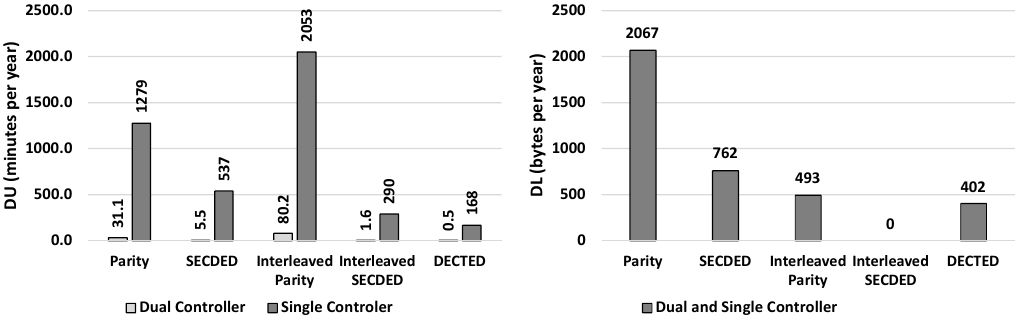}
        \label{fig:dudl-websearch1}
    }
    \subfigure[Websearch\_2 Workload]
	{
		\includegraphics[width=0.5\textwidth]{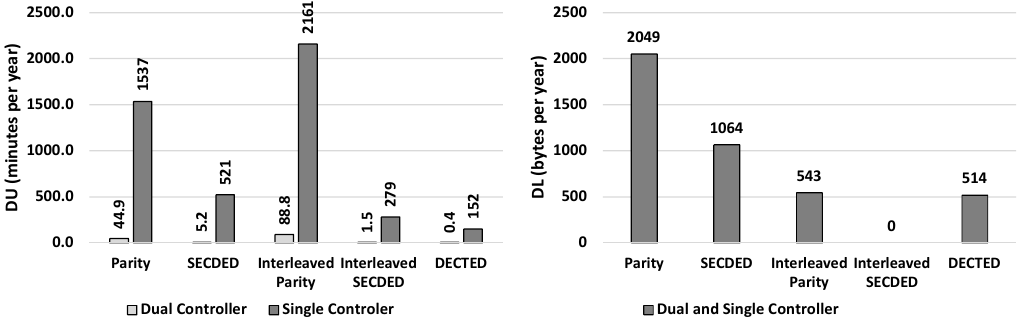}
        \label{fig:dudl-websearch2}
    }
        \subfigure[Websearch\_3 Workload]
	{
		\includegraphics[width=0.5\textwidth]{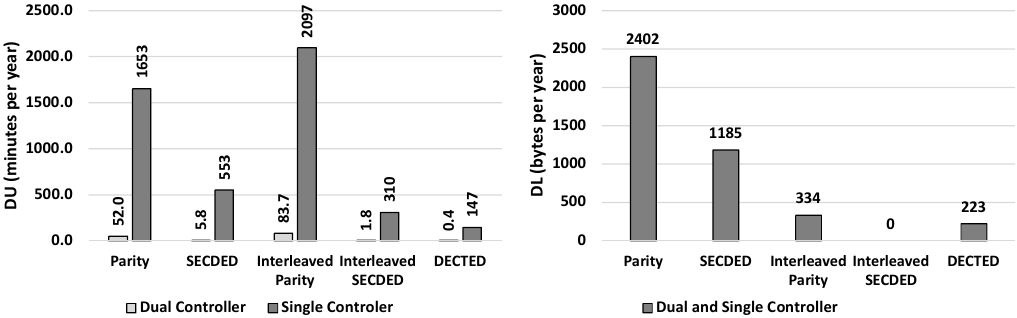}
        \label{fig:dudl-websearch3}
    }   
    \caption{DU (minutes) and DL (bytes) in one year mission (assuming 1000 SEU/year) for single and dual controller with different cache protection schemes (simulation configuration appeared in Table~\ref{tab:configuration}).}
    \label{fig:normalizeddudl}
\end{figure}

{Fig.~\ref{fig:normalizeddudl} shows the DU (minutes per year) and DL (bytes per year) values for different cache protection schemes (the same protection is assumed for both tag and data fields). These values are obtained using our fault injection experiments under Financial\_1, Financial\_2, Websearch\_1, Websearch\_2, and Websearch\_3 workloads, for both single and dual controller architectures. Note that both single and dual controller architectures perform similar in terms of DL (as discussed in Section~\ref{sec:systemlevel-dual}). The reason is that unlike \emph{Duplication With Comparison} (DWC) architecture in which the execution is duplicated on two redundant processing units and the output is verified by comparing two redundant results, in dual controller architecture, two controllers are independently performing different tasks (when both controllers are operational). Hence, dual controller architecture is not designed to detect/correct DL happening in individual controllers. However, this architecture can prevent DU incidence upon the failure of one controller, by redirecting the tasks of the failed controller to the operational one. 

As the results show, none of the examined cache protections schemes can outperform the others in terms of both DU and DL. For example, interleaved SECDED shows the lowest DL (zero), as it can detect all errors in {our} experiments while it shows higher DU compared to DECTED in both Financial\_1 and Websearch\_1 workloads. 
{Both interleaved SECDED and DECTED protections perform the same in the case of 3-bit MBUs (detect) when considering $M\times 1$ error bit pattern (as described in Section~\ref{sec:error bit pattern}).}
Greater DU of interleaved SECDED {compared to DECTED} can be described by the fact that 4-bit MBUs, resulting in DL in the case of DECTED, are detected by interleaved SECDED and result in DU. Meanwhile, both linear parity and interleaved parity schemes show a relatively high DU among all schemes. This {observation is} described by the fact that parity has a relatively high detection capability, but zero correction. Hence, DUEs will result in a high rate of controller reboot, resulting in DU. 

An important observation {is that} the ranking of protection schemes in both DU and DL is the same as their $SSVF^{DU}$ and $SSVF^{DL}$ ranking, showing that $SSVF$ can be an effective representative for comparing different protection schemes in terms of data unavailability and data loss. Regarding the relationship between DU and $SSVF^{DU}$, we can observe an analogous shape of diagram. {This observation is described by the fact that} the reported DU (in terms of minutes) is simply number of DU incidences multiplied by reboot time, while $SSVF^{DU}$ is formulated as the number DU incidences divided by the number of fault injections. The relationship between DL (in terms of bytes) and $SSVF^{DL}$, however, is more complicated. DL caused by {soft-errors} on tag and data field do not have the same magnitude (tag soft-error results in the whole line, 64 bytes, loss while soft-errors on the data field pollute \emph{only} one data word, 8 bytes, or two adjacent data words). Meanwhile in calculating DL {we} also collect the DL propagation statistics (by checking data re-use and propagation of polluted data to other memory hierarchies) that is not included in $SSVF^{DL}$ calculation. Consequently, DL values of different protection schemes do not relate exactly the same as $SSVF^{DL}$. }


\begin{figure}
    \centering
 
	\subfigure[DU] 
	{
        \includegraphics[width=0.4\textwidth]{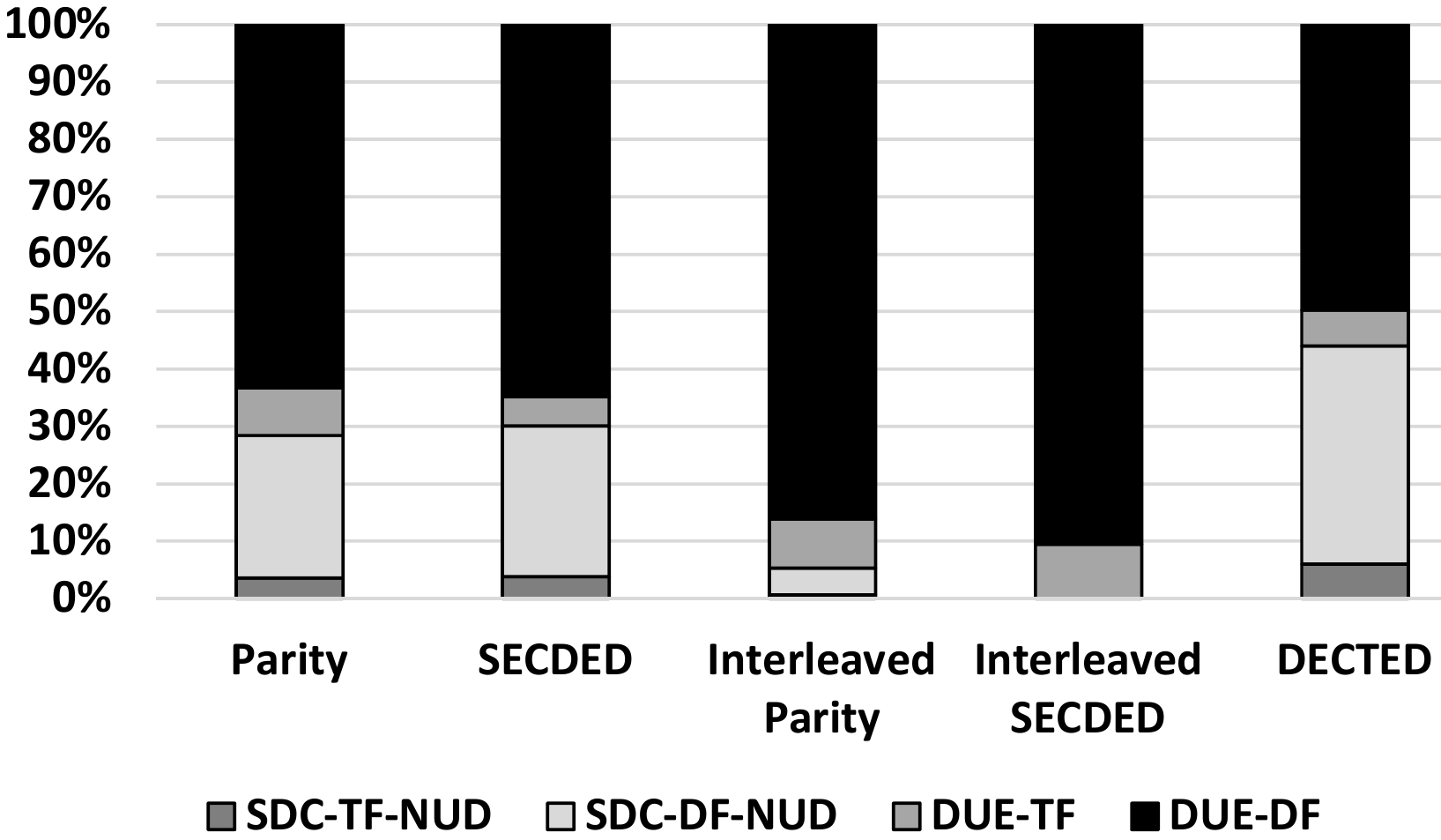}
        \label{dudlbreakdown-du}
    }
    
    \subfigure[DL]
	{
		\includegraphics[width=0.4\textwidth]{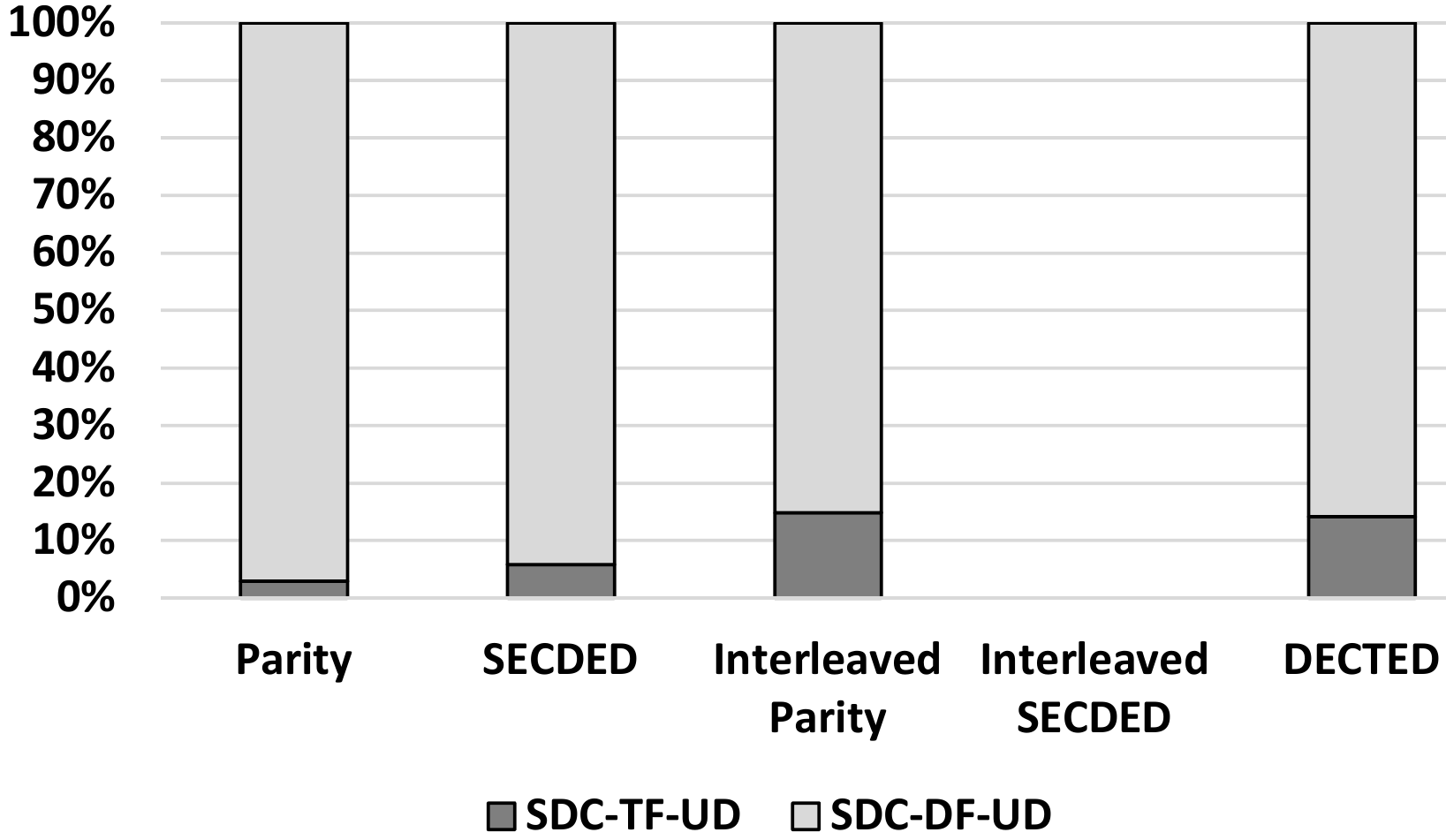}
        \label{dudlbreakdown-dl}
    }
   
    \caption{{DU/DL breakdown for different cache protection schemes (Financial\_1 workload). DUE-TF is reported as the aggregation of DUE-TF on user data (DUE-TF-UD) and DUE-TF on non-user data (DUE-TF-NUD). Similarly, DUE-DF is reported as the aggregation of DUE-DF on user data (DUE-DF-UD) and DUE-DF on non-user data (DUE-DF-NUD).}}
    \label{fig:dudlbreakdown}
\end{figure}

{Fig.~\ref{fig:dudlbreakdown} shows the breakdown of DU (hours) and DL (bytes) at the storage level, showing the fraction of DU and DL {caused by} 
DUE and SDC (for Financial\_1 workload). 
DL chart shows that {in all protection schemes,} {the most data loss} is caused by SDC on the data field of user data. 
Investigating the DU breakdown shows that DUE on data field is the major source of DU in all protection schemes. 
The second source of DU is SDC on data field of non-user cache blocks. Hence, here we also can conclude that soft-errors on the cache data field are the major source of DU.}

\begin{figure*}
    \centering
 \subfigure[Share of MBU in total DU/DL for different cache protection schemes] 
	{
        \includegraphics[width=0.7\textwidth]{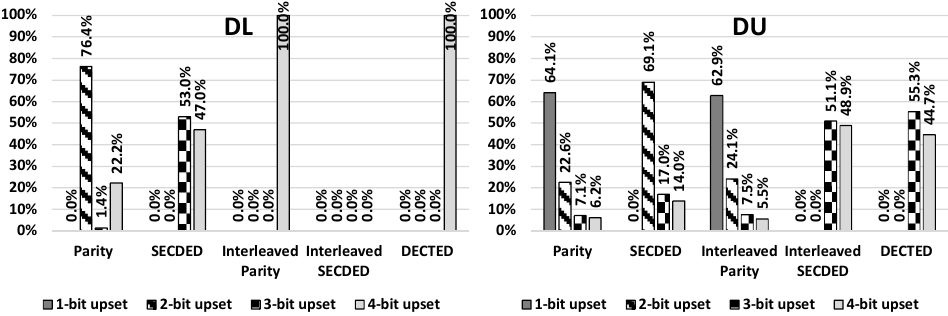}
    \label{fig:mbududl}
   } 
    
 \subfigure[Number of DU and DL incidences by different types of MBU, normalized to the number of injected faults from each type of MBU] 
	{
        \includegraphics[width=0.7\textwidth]{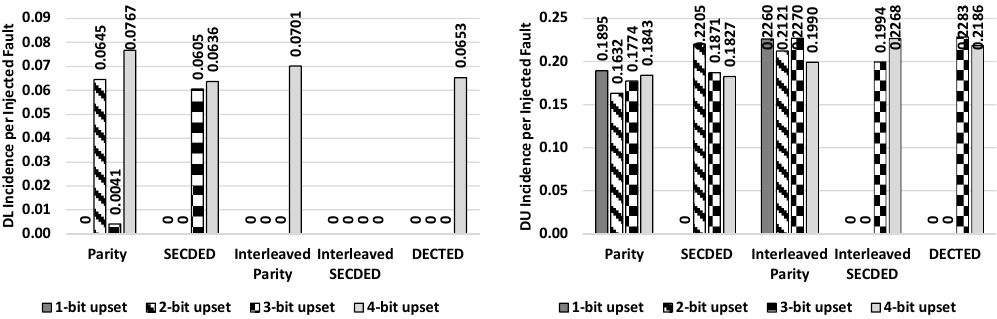}
    \label{fig:mbududl-normalized}
   }

    \caption{Effect of different MBU types on DU/DL of data storage system (Financial\_1 workload).}
    \label{fig:effect-mbududl}

\end{figure*}

{Fig.~\ref{fig:mbududl} shows the fraction of DU and DL incidences caused by single and multiple bit upsets.
A notable point in Fig.~\ref{fig:mbududl} is non-zero DL caused by 3-bit MBUs when having parity protection. The first impression is that parity {can detect} 3-bit MBUs and no DL is expected by this type of MBU. However, there is a possibility that 3-bit MBU targets two adjacent words (MCU). In that case, one word is polluted by a single bit flip while its {adjacent} word is polluted by two bit flips which {is} not detectable by parity code, resulting in DL.}

 {Fig.~\ref{fig:mbududl-normalized} shows the 
 number of DU and DL incidences by different types of MBU, normalized to the number of injected faults from each type of MBU.
{As the results show, in the case of parity code, 2-bit and 4-bit MBUs have almost the same chance to result in DL. This observation was predictable as the parity code fails to detect even number of bit flips, including both 2-bit and 4-bit MBUs. However, 2-bit MBUs have a slightly less chance than 4-bit MBUs to result in DL, described by the cases in which 2-bit MBU targets two adjacent words (MCU). In such cases, each individual word is polluted by a single bit flip that is detected using parity code.}
 The same happens in the case of 3-bit and 4-bit MBUs for SECDED protection.

 {In the} case of interleaved parity, 1-bit, 2-bit, and 3-bit upsets have almost the same chance leading to DU, as they are all detectable. We also observe that 4-bit MBUs, resulting in SDC in the case of interleaved parity, also have a high chance becoming DU. This observation can be described by the fact that SDCs targeting non-user data have also the chance to result in DU. The same happens about 2-bit and 4-bit MBUs when using parity protection, 3-bit and 4-bit MBUs when using SECDED protection, and 4-bit MBUs when using DECTED protection. In the case of interleaved SECDED, we observe that detectable errors caused by 3-bit and 4-bit MBUs have almost the same chance to result in DU. }

\subsection{{DL Propagation}}
\begin{figure}
    \centering

        \includegraphics[width=0.45\textwidth]{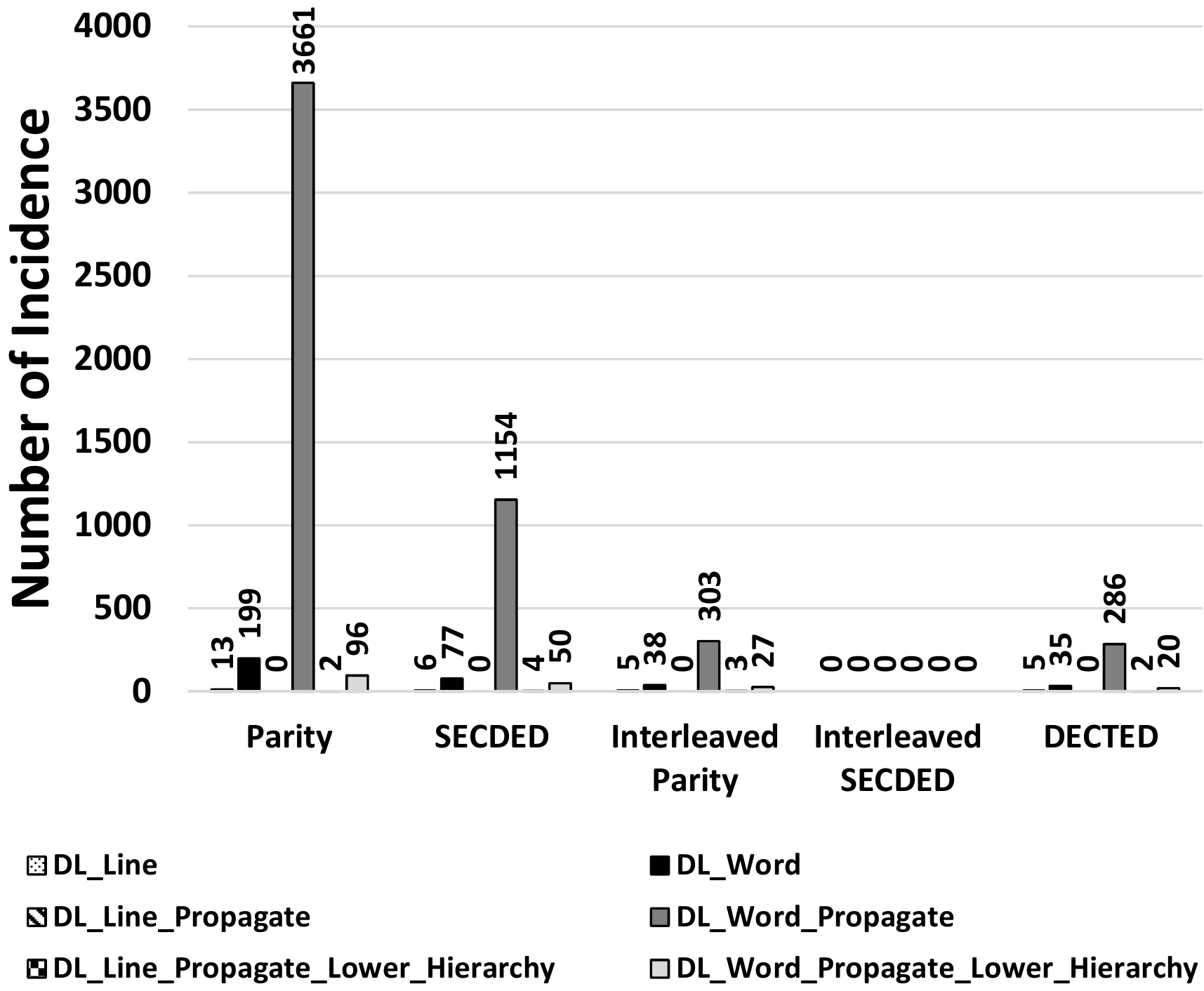}
    \caption{The effect of DL propagation (Financial\_1 workload)}
    \label{fig:dl-propagation}

\end{figure}

{We consider the effect of DL propagation, as noted in Section~\ref{sec:fault-tracking}. 
To clarify the contribution of DL propagation in total DL, Fig.~\ref{fig:dl-propagation} shows the number of different DL incidences for 10,000 injected faults (Financial\_1 workload). DL\_Line shows the initial DL caused by SDCs targeting tag field of user cache blocks, resulting in the loss of entire cache line. Similarly, DL\_Word shows the initial DL caused by SDCs targeting data field of user cache blocks, resulting in the loss of one (or in the most intense scenario, two) data word. DL\_Line\_Propagate is named after the case the DL in the entire line (caused by soft-error on tag field) is propagated by accessing the entire line. This case never happens, as the access resolution to cache blocks is one data word. The entire line is updated just in the case of line \emph{Update} and \emph{Evict} {operations}. 
DL\_Word\_Propagate refers to the DL propagation caused by reusing (reading) the faulty data. Note despite the fault {targets either of} tag field or data field, reading a faulty word is recognized as a DL\_Word\_Propagate. DL\_Line\_Propagate\_Lower\_Hierarchy is named after the case a faulty line (caused by {soft-error on the tag field}) is evicted, while it has LINE\_MODIFIED (or MESI\_MODIFIED in the case of coherent cache) status. In {that} case, the faulty line should be written back to the lower memory hierarchy. Hence, the DL is propagated to the lower hierarchy. Finally, DL\_Word\_Propagate\_Lower\_Hierarchy stands for the case a cache line holding a faulty word (caused by soft-error on {the}  data field) is evicted, while it has LINE\_MODIFIED (or MESI\_MODIFIED in the case of coherent cache) status. In this case, a one-word DL is propagated to the lower memory hierarchy. 
As the results show, DL\_Word\_Propagate has the most contribution in total DL for all protection schemes. The results also show that the effect of DL propagation is one order of magnitude greater than initial DL.}

\subsection{{Masking Effect}}
\begin{figure}
    \centering

        \includegraphics[width=0.45\textwidth]{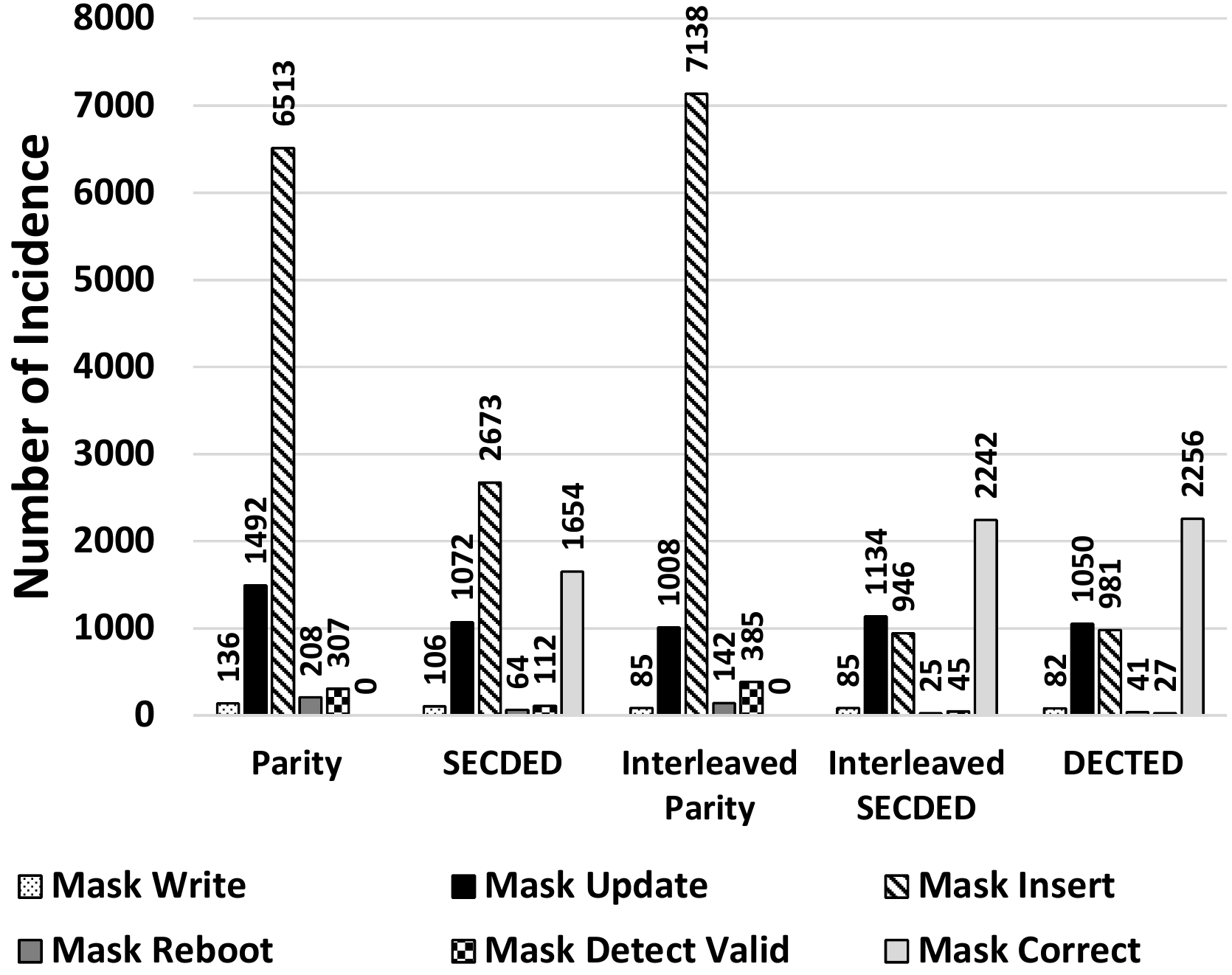}
    \caption{The effect of fault masking (Financial\_1 workload)}
    \label{fig:masking-effect}

\end{figure}

{{In the fault injection experiments, we} track and report the error masking cases, as noted in Section~\ref{sec:fault-tracking}. Fig.~\ref{fig:masking-effect} shows the number of error masking incidences observed in 10,000 fault injections for Financial\_1 workload. Mask\_Write refers to the case an error is masked by overwriting the faulty data. Mask\_Update is named after the case a cache line is updated and the error (in data field) is totally masked. Mask\_Insert refers to the case a new line is inserted and the faulty cache line is evicted (in this case, the error is masked if the cache line does not have LINE\_MODIFIED status. In the case of LINE\_MODIFIED statues, the error is propagated to the lower hierarchy). Mask\_Reboot refers to the faults that cause controller reboot. In that case, the faulty data is removed after the controller new startup. {Mask\_Detect\_Valid} refers to the case an error in a valid cache line (i.e., a clean cache line whose copy exists in {either of} lower memory hierarchies) is detected. In that case, the correct data is obtained from lower memory hierarchy and the error is masked at no DU/DL cost. The final case, Mask\_Correct, refers to the case an error is correctable. In that case, the error is corrected at no DU/DL cost.  
As the results show, in parity, SECDED, and interleaved parity protections, Mask\_Insert has the most contribution in error masking. Please note that these stats are presented for all injected faults, including the faults injected on invalid cache lines. In Interleaved SECDED and DECTED protections which have the greatest correction capability (both can correct up to 2-bit MBUs), we observe Mask\_Correction has the most contribution in error masking. }

\subsection{Impact of Workloads}
\begin{figure}
    \centering
 
	\subfigure[Financial\_1 Workload] 
	{
        \includegraphics[width=0.4\textwidth]{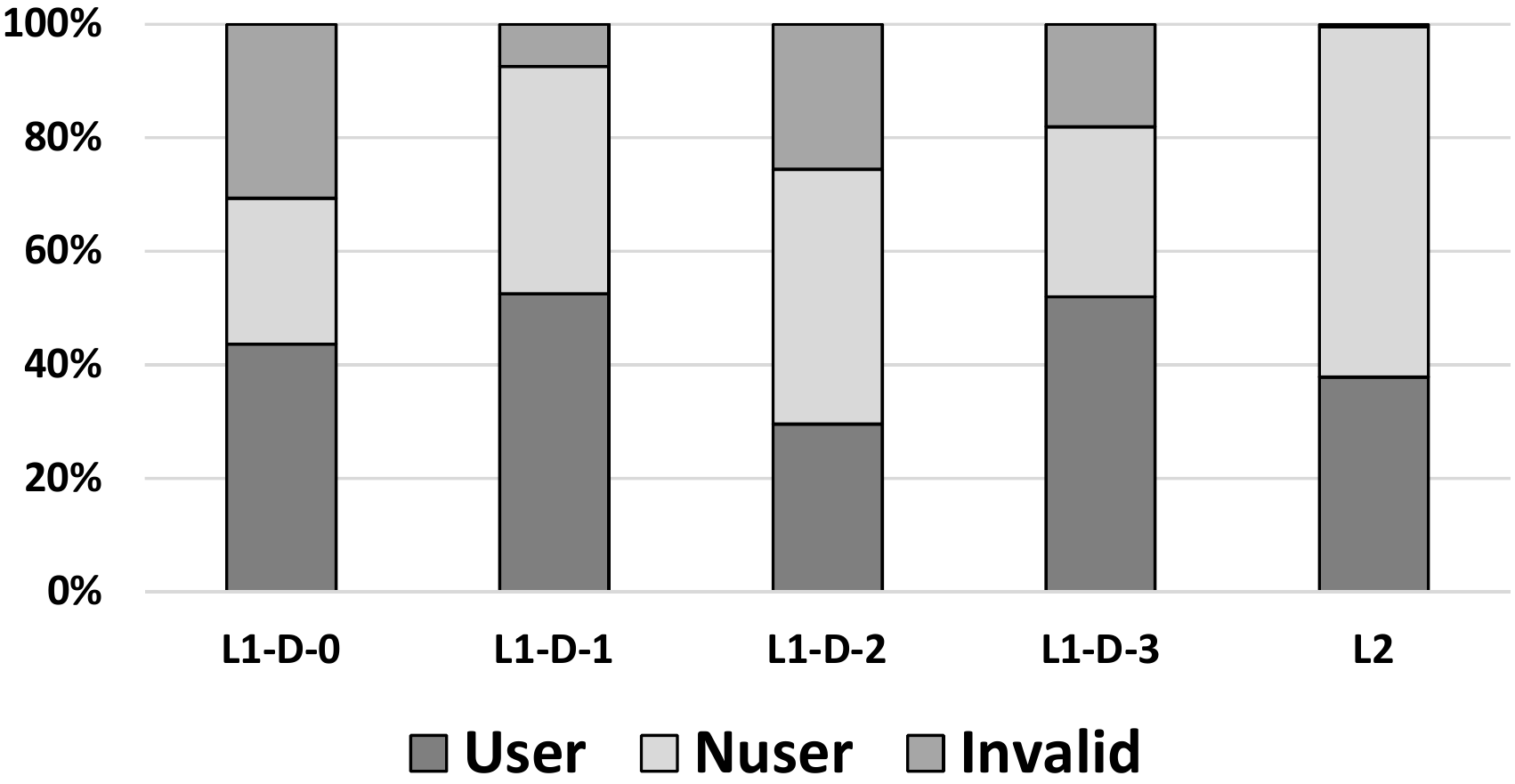}
        \label{fig:financial-userdata}
    }
    
    \subfigure[Websearch\_1 Workload]
	{
		\includegraphics[width=0.4\textwidth]{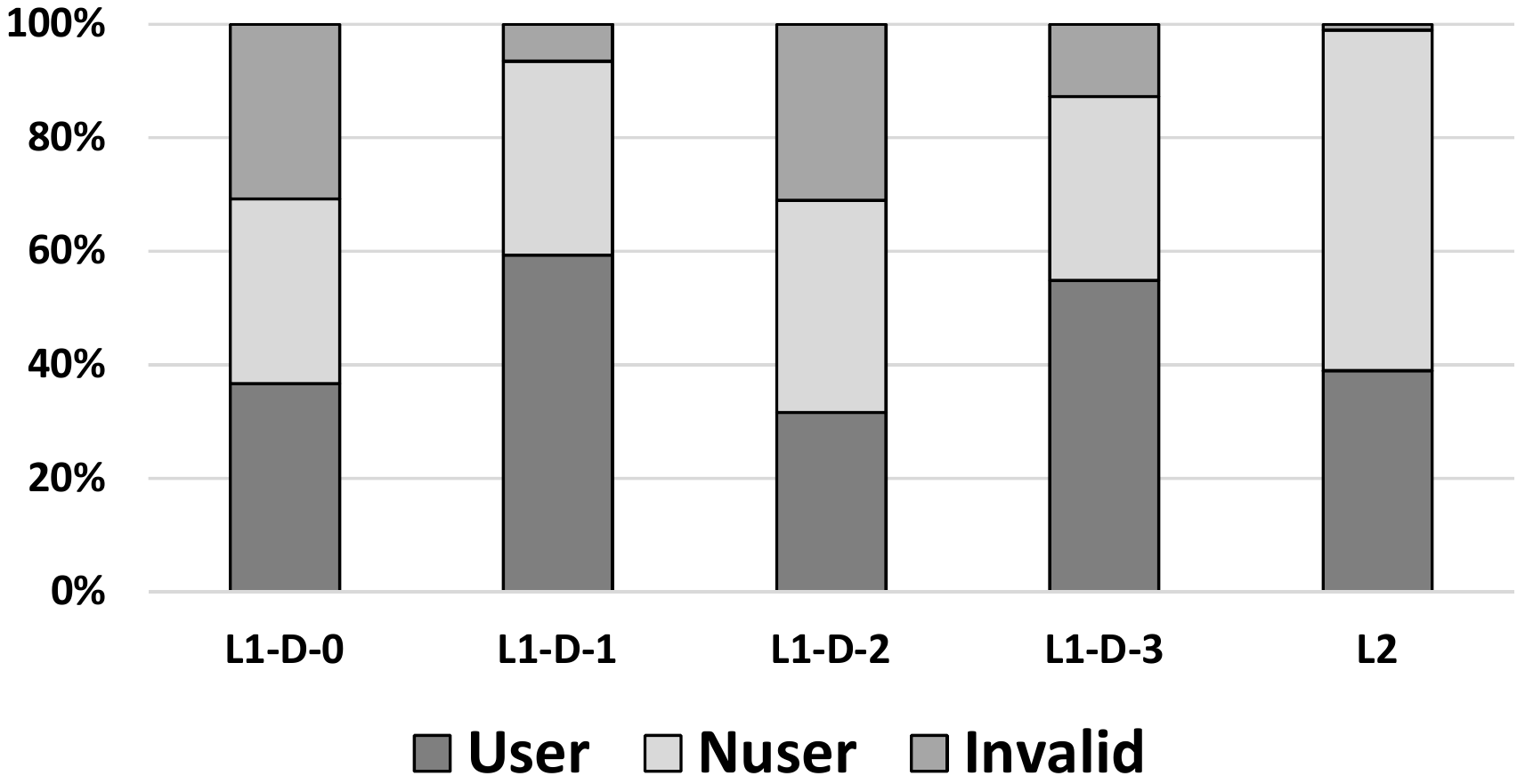}
        \label{fig:websearch-userdata}
    }
   
    \caption{The average fraction of cache memory occupied by end-user data (\emph{User}), non-user data (\emph{Nuser}), and invalid data (\emph{Invalid}) in storage controller processors.}
    \label{fig:userdata-financial-websearch}
\end{figure}

In this section, we investigate the effect of workload (discussed in Section~\ref{sec:examinedworkloads}) on data unavailability and data loss. 
Fig~\ref{fig:userdata-financial-websearch} shows the average fraction of cache memory occupied by end-user data, non-user data, and {invalid data}. 
The part of cache memory that is occupied by end-user data is susceptible to data loss at storage system level 
(in the case a soft error results in SDC in the cache memory).
{The results show that when running Financial\_1 workload, respectively 44\% and 37\% of L1 and L2 cache memory is (on average) occupied by end-user data. For Websearch\_1 workload, respectively 46\% and 39\% of L1 and L2 cache memory is occupied by end-user data. 
In some periods of mission time, usually when the storage system is handling a burst of requests, we observe that more than 80\% of cache memory is occupied by the end-user data, in both Financial\_1 and Websearch\_1 workloads.}

\begin{figure*}
    \centering
         \includegraphics[width=1\textwidth]{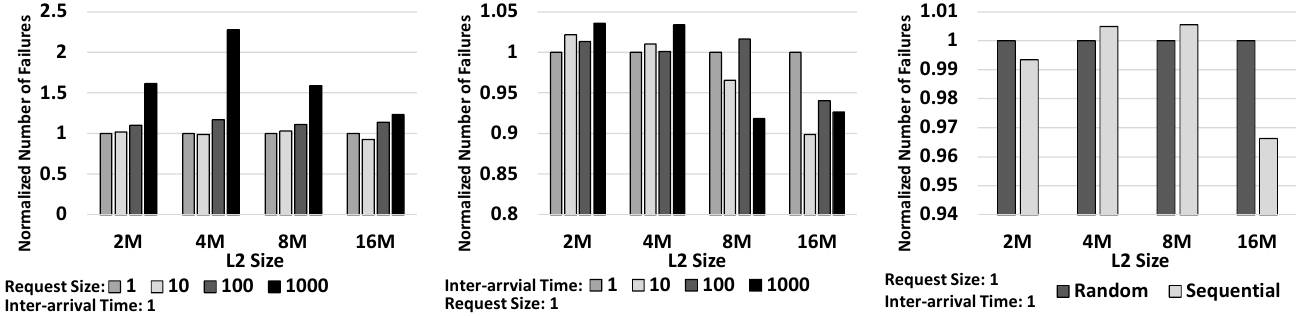}
    \caption{{Effect of request size (KB), inter-arrival time (micro seconds), and randomness, on the number of failures (aggregation of DU and DL incidences). Request size results (the left-most chart) are obtained by considering \emph{Inter-arrival Time = $1\mu s$} and sequential workload, and are normalized to $1KB$ request size. Inter-arrival time results (the middle chart) are obtained by considering \emph{Request Size = 1KB} and sequential workload, and are normalized to $1\mu s$. Randomness results (the right-most chart) are obtained by considering \emph{Inter-arrival Time = $1\mu s$} and \emph{Request Size = 1KB}, and are normalized to Random. The experiments are conducted for different sizes of L2 cache (from 2MB to 16MB) by considering single storage controller. In the experiments we consider \emph{No Protection} for the cache memory and 10,000 faults are injected per configuration.}}
    \label{fig:normalized-syntheitc}

\end{figure*}

{Fig.~\ref{fig:normalized-syntheitc} shows the effect of request size (KB), inter-arrival time (micro seconds), and randomness, on the number of failures (aggregation of DU and DL incidences).
As the figure shows, the number of failures is directly proportional to the request size (the left-most chart). By increasing the request size from 1KB to 1000KB, the failure rate is increased by up to 2.3 times.  
The impact of inter-arrival time, however, is not significant (the middle chart). Increasing inter-arrival time results in up to 10\% variation in number of failures. We also observe that number of failures is not a monotonic function of inter-arrival time. For 2MB and 4MB L2 size, the number of failures is slightly ascending with inter-arrival time, while for 8MB and 16M L2 size, number of failures is a descending function of inter-arrival time.
The results also show that randomness has a negligible effect on the number of failures (the right-most chart). 
In $L2=4MB$ and $L2=8MB$ configurations, sequential requests show $0.4\%$ and $0.5\%$ greater number of failures compared to random requests, 
while in $L2=16MB$ configuration, random requests show $4\%$ greater number of failures compared to sequential requests.  
Hence we can conclude that the failure rate is almost independent of request randomness.
}

\subsection{{Impact of $P_{NotManifest}$ and $P_{OS_{DL}}$ }}
{In this section, we investigate the effect of $P_{NotManifest}$ and $P_{OS_{DL}}$ (both defined in Section~\ref{sec:read-non-dirty}. Please note that in {the experiments,} we do not capture $P_{NotManifest}$, as we assume once an SDC on non-user data is activated it will result in OS/application malfunction. 
Our simulations also do not capture $P_{OS_{DL}}$ which is the probability of SDC on {non-user} data leading to DL, due to OS/applications malfunction.
Here we use the empirical data obtained by Gu et al.~\cite{gu2003characterization} that reports both $P_{NotManifest}$ and $P_{OS_{DL}}$ by {injecting fault} on important modules of Linux kernel. 
In summary, Gu et al. report 30.4\% of SDCs injected to four most important subsystems of Linux OS (representing more than 95\% of kernel usage) are not manifested ($P_{NotManifest}$ is equal to 30.4\%). 
This study also shows that out of 35000 faults injected to Linux subsystems, 9 cases result in filesystem crashes. Despite all observed crashes in this study target OS addressing space, there is a possibility that they also target user space in some cases and result in user data loss ({hence,} $P_{OS_{DL}}$ is equal to 0.00025). }

\begin{figure}
    \centering
 
        \includegraphics[width=0.5\textwidth]{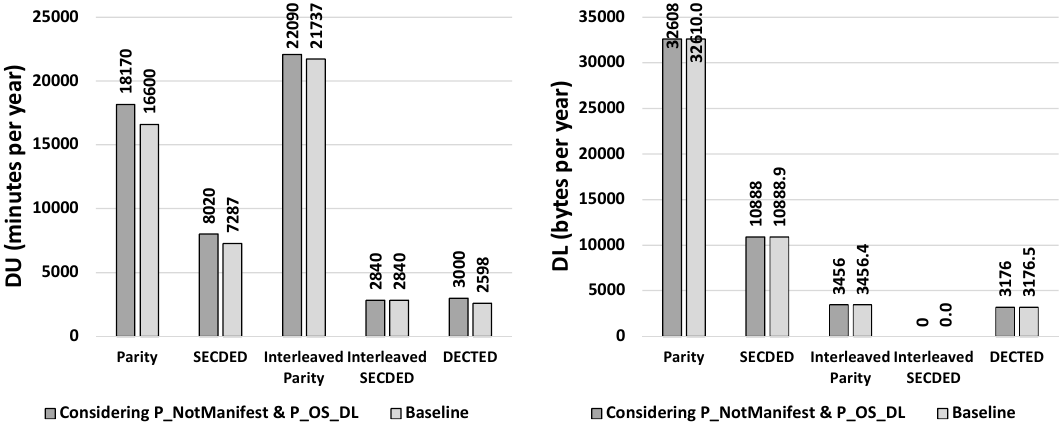}
    \caption{Effect of $P_{NotManifest}$ and $P_{OS_{DL}}$ (Financial\_1 workload)}
    \label{fig:du-dl-os-dl}

\end{figure}

{Fig.~\ref{fig:du-dl-os-dl} compares our baseline results with results gathered by considering {the effect of} $P_{NotManifest}$ and $P_{OS_{DL}}$ (obtained by~\cite{gu2003characterization}).  
As the results show, considering $P_{OS_{DL}}$ has a negligible effect on {total} DL ({it results} in less than 0.02\% DL increase in all protection schemes). 
However, the effect of $P_{NotManifest}$ is more considerable, as it decreases DU by up to 13\% (in the case of DECTED). Hence, we can conclude that ignoring $P_{NotManifest}$ may result in DU overestimation in our experiments.
}

\subsection{{Impact of Soft-Errors in Cache Control Logic}}
{In this section, we investigate the effect of soft-errors in cache control logic. The errors of cache control logic are not detectable/correctable by ECC and affect the entire cache line, rather than a single data word. 
To this end, we estimate the area of cache control logic using CACTI 7.0 tool~\cite{cacti7}. We also use the data from Shivakumar et al.~\cite{shivakumar2002modeling} for \emph{Soft-Error Rate} (SER) on combinational logic, resulting from high-energy Neutrons. This study does not consider the effect of logical masking and simply reports SER as a function of number of combinational logic chains (logics using 2-input NAND gates with \emph{Fan-Out 4}, FO4) and the length of logic chain (i.e. the working frequency). Using the area of cache control logic, we estimate the number of logic chains and evaluate the SER of cache controller, in terms of \emph{Failure in Time} (FIT)\footnote{{The number of failures per $10^{9}$ hours of operation.}} as summarized in Table~\ref{tab:ser-cache-controller}.
 }

\begin{table}
\centering
\caption{{Number of FO4 logics with length of 12 and SER of cache memory controller}}
\vspace{-0.6cm}
\begin{center}
\begin{adjustbox}{width=\textwidth/2,totalheight=\textheight,keepaspectratio}
    \begin{tabular}{ | c | c | c | c |}
    \hline
& L1 & L2 & Total (4-Core, dedicated IL1/DL1, shared L2) \\ \hline
Number of 12-FO4 Logics & 125232 & 756499 & 1758362 \\ \hline
SER (FIT/Controller) & $5.00$ & $30.25$ & 70.33 \\ \hline

    \end{tabular}
    \end{adjustbox}
\end{center}

\label{tab:ser-cache-controller}
\end{table}

\begin{figure}
    \centering
 
        \includegraphics[width=0.5\textwidth]{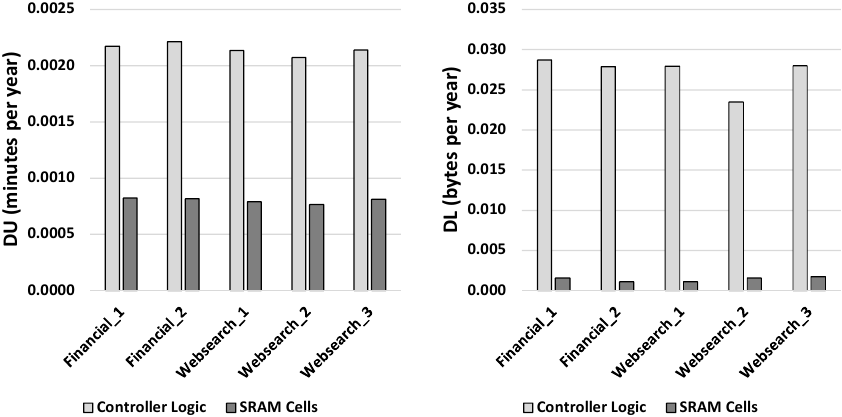}
    \caption{Expected annual DU and DL caused by soft-errors in cache controller logic and SRAM cells (considering SECDED protection) for different real workloads}
    \label{fig:du-dl-controller}

\end{figure}

{Using the SER estimated for the entire cache controller (Table~\ref{tab:ser-cache-controller}), we inject faults to the processor cache controller assuming that the errors injected to the control logic are neither correctable nor detectable by the cache ECC. Hence, each fault in the control logic is interpreted as an undetectable tag fault. We perform 10,000 fault injection experiments and normalize {DU/DL results} to the number of soft-errors expected in one year mission time. Using FIT/Controller (Table~\ref{tab:ser-cache-controller}), the annual expected number of soft-errors per cache controller is $0.000616$. Accordingly, the expected DU/DL per year for different real benchmarks is shown in Fig.~\ref{fig:du-dl-controller}.
This figure also shows the expected DU/DL per year caused by soft-errors on cache SRAM cells (obtained by using FIT/SRAM reported by Shivakumar et al.~\cite{shivakumar2002modeling}), for SECDED cache protection.
As the results show, the expected DU caused by soft-errors in the controller logic is more than two times greater than SRAM cells. In the case of DL, the difference is even more, as we observe DL caused by soft-errors in controller logic is one order of magnitude greater than SRAM cells. The significant impact on DL is described by the fact that all soft-errors in the controller logic go undetected, resulting DL if they target end-user data.
This observation shows that cache controller reliability has a great importance in DU/DL prevention, seeking for more detailed studies and investigations.     
}

\section{{Conclusion and Discussion}}
\label{sec:conclusion}
{In this paper, we modeled the storage system level effects of soft errors occurring in the controller cache memory. We set up our framework by first implementing the major functions of storage controller, running on a full stack of Linux kernel, and then developing a framework to perform fault injection experiments using a full system simulator. We proposed a new metric, called SSVF, defined as the probability that a soft error results in DU/DL at the storage level, as an alternative to AVF that cannot directly represent the DU/DL of a specific storage design.
We can conclude the main findings of this work as follows:}
\begin{itemize}
\item
{Comparing AVF results with SSVF results shows that AVF does not correctly represent the chance of DU and DL in data storage systems.}
\item
{Cache protection schemes with greater detection capability always experience lower DL. However, improving error detection may have an ascending effect on DU, as the controller reboots itself upon a detectable unrecoverable error to prevent data loss.}
\item
{Interleaved SECDED is the most reliable protection scheme, leading to lowest amount of DL, while DECTED is the most efficient protection schemes in terms of availability.}
\item
{Tag fields targeted by soft-errors are more vulnerable to both DU and DL compared to data fields.}
\item
{DUEs on the cache data field are the major cause of DU in all protection schemes, while SDCs on the data field of user cache blocks contribute to the most DL.}
\item
{The DL propagated by reusing the faulty data has the most contribution in total DL reported, while the effect of DL propagation is one order of magnitude greater than the initial DL.}
\item
{We observed different sources of error masking in our experiments. For parity, SECDED, and interleaved parity protection, inserting a new cache line (cache evict) is the major source of masking, while for interleaved SECDED and DECTED protections, the error correction has the most contribution in error masking.}
\item
{By increasing the average request size, the storage failure rate considerably increases, while request inter-arrival time and randomness do not have significant effect on the failure rate.}
\item
{While this work is mainly focused on soft-errors in cache SRAM cells, our approximations on the effect of soft-errors in cache controller logic have been so motivative. We observe the expected DU caused by soft-errors in the controller logic is more than twice greater than SRAM cells. In the case of DL, we approximate DL caused by soft-errors in the controller logic is one order of magnitude greater than SRAM cells. This observation is described by the fact that soft-errors in the controller logic go undetected, resulting in DL if they target end-user data.}
\end{itemize}

{In the future work, we will investigate the effect of software robustness and software-level protections on the reliability and availability of storage controllers.}



%



\ifCLASSOPTIONcaptionsoff
  \newpage
\fi



%

\bibliographystyle{IEEEtran}
\bibliography{bare_jrnl}

%

\begin{IEEEbiography}[{\includegraphics[width=1in,height=1in,clip]{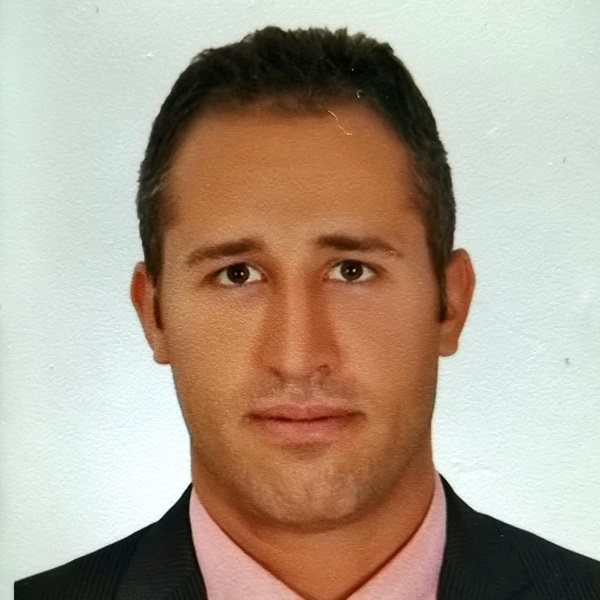}}]%
{Mostafa Kishani}
received the B.S. degree in computer engineering from Ferdowsi University of Mashhad, Mashhad, Iran, in 2008, and M.Sc. degree in computer 
Engineering from Amirkabir University of Technology (AUT), Tehran, Iran, in 2010. 
He is currently a PhD student of computer engineering in the Sharif University of Technology (SUT), Tehran, Iran, since 2012.
He was a hardware engineer in Iranian Space Research Center (ISRC) from 2010 to 2012.
He was also a member of Institute for Research in Fundamental Sciences (IPM) Memocode team in 2010.
From September 2015 to April 2016 he was a research assistant in Computer Science and Engineering department of the Chinese University of 
Hong Kong (CUHK), Hong Kong.
He was also a research associate in the Hong Kong Polytechnic University (PolyU), Hong Kong, from April 2016 to February 2017.

\end{IEEEbiography}

\begin{IEEEbiography}[{\includegraphics[width=1in,height=1.5in,clip]{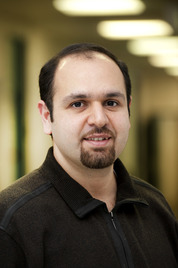}}]%
{Mehdi Tahoori}
Mehdi Tahoori is a full professor and Chair of Dependable Nano-Computing (CDNC) at the Institute of Computer Science \& Engineering (ITEC), Department of Computer Science, Karlsruhe Institute of Technology (KIT), Germany. He received his PhD and M.S. degrees in Electrical Engineering from Stanford University in 2003 and 2002, respectively, and a B.S. in Computer Engineering from Sharif University of Technology in Iran, in 2000. In 2003, he joined the Electrical and Computer Engineering Department at the Northeastern University as an assistant professor where he promoted to the rank of associate professor with tenure in 2009. From August to December 2015, he was a visiting professor at VLSI Design and Education Center (VDEC), University of Tokyo, Japan. From 2002 to 2003, he was a Research Scientist with Fujitsu Laboratories of America, Sunnyvale, CA, in the area of advanced computer-aided research, engaged in reliability issues in deep-submicrometer mixed-signal very large-scale integration (VLSI) designs.
He holds several pending and granted U.S. and international patents. He has authored over 250 publications in major journals and conference proceedings on a wide range of topics, from dependable computing and emerging nanotechnologies to system biology. His current research interests include nanocomputing, reliable computing, VLSI testing, reconfigurable computing, emerging nanotechnologies, and systems biology.
Prof. Tahoori was a recipient of the National Science Foundation Early Faculty Development (CAREER) Award. He has bee a program committee member, organizing committee member, track and topic chair, as well as workshop, panel, and special session organizer of various conferences and symposia in the areas of VLSI design automation, testing, reliability, and emerging nanotechnologies, such as ITC, VTS, DAC, ICCAD, DATE, ETS, ICCD, ASP-DAC, GLSVLSI, and VLSI Design. He is currently an associate editor for IEEE Design and Test Magazine (D\&T), coordinating editor for Springer Journal of Electronic Testing (JETTA), associate editor of VLSI Integration Journal, and associate editor of IET Computers and Digital Techniques. He was an associate editor of ACM Journal of Emerging Technologies for Computing. He received a number of best paper nominations and awards at various conferences and journals, including ICCAD 2015 and TODAES 2017. He is the Chair of the ACM SIGDA Technical Committee on Test and Reliability.

\end{IEEEbiography}

\begin{IEEEbiography}[{\includegraphics[width=1in,height=1.25in,clip]{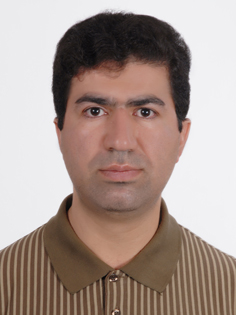}}]%
{Hossein Asadi}
(M'08, SM'14) received the B.Sc. and M.Sc. degrees in computer engineering from the SUT, Tehran, Iran, in 2000 and 2002, respectively, and the Ph.D. degree in electrical and computer engineering from Northeastern University, Boston, MA, USA, in 2007. 

He was with EMC Corporation, Hopkinton, MA, USA, as a Research Scientist and Senior Hardware Engineer, from 2006 to 2009. From 2002 to 2003, he was a member of the Dependable Systems Laboratory, SUT, where he researched hardware verification techniques. From 2001 to 2002, he was a member of the Sharif Rescue Robots Group. He has been with the Department of Computer Engineering, SUT, since 2009, where he is currently a tenured Associate Professor. He is the Founder and Director of {the DSN Laboratory}, Director of Sharif \emph{High-Performance Computing} (HPC) Center, the Director of Sharif \emph{Information and {Communications} Technology Center} (ICTC), and the President of Sharif ICT Innovation Center. He spent three months in the summer 2015 as a Visiting Professor at the School of Computer and Communication Sciences at the Ecole Poly-technique Federele de Lausanne (EPFL). He is also the co-founder of HPDS corp., designing and fabricating midrange and high-end data storage systems. He has authored and co-authored more than eighty technical papers in reputed journals and conference proceedings. His current research interests include data storage systems and networks, solid-state drives, operating system support for I/O and memory management, and reconfigurable and dependable computing.

Dr. Asadi was a recipient of the Technical Award for the Best Robot Design from the International RoboCup Rescue Competition, organized by AAAI and RoboCup, a recipient of Best Paper Award at the 15th CSI {International} Symposium on \emph{Computer Architecture \& Digital Systems} (CADS), the Distinguished Lecturer Award from SUT in 2010, the Distinguished Researcher Award and the Distinguished Research Institute Award from SUT in 2016, and the Distinguished Technology Award from SUT in 2017. He is also recipient of Extraordinary Ability in Science visa from US Citizenship and Immigration Services in 2008. He has also served as the publication chair of several national and international conferences including CNDS2013, AISP2013, and CSSE2013 during the past four years. Most recently, he has served as a Guest Editor of IEEE Transactions on Computers, an Associate Editor of Microelectronics Reliability, a Program Co-Chair of CADS2015, and the Program Chair of CSI National Computer Conference (CSICC2017). 
\end{IEEEbiography}




\newpage

\appendices
\section{{Fault Tracking}}
\label{appendix:dl propagation}
\vspace{-0.1cm}
{In the following, we detail the cache structure and fault masking/propagation mechanism in both shared cache (both L2 and IL1/DL1 caches when having single-core architecture) and MESI coherent cache (IL1/DL1 cache when having multi-core architecture). }

\vspace{-0.3cm}
\subsection{{Cache States in MARSSx86}}
\subsubsection{{Shared Cache Line States}}
{In the shared cache architecture, we have the following cache status:}
\begin{itemize}
	\item
	{LINE\_NOT\_VALID: The cache line is not valid. In this case the SDC is masked despite it is on data field or tag field.}  
	\item
	{LINE\_VALID: The cache line is valid and clean. Cache insert (evicting the old line) in this case does not need write-back to lower memory hierarchy. So the SDC is not propagated to the lower memory hierarchy.} 
	\item
	{LINE\_MODIFIED: The cache line is modified. In this case, the cache line contains the only valid copy of data and cache insert mandates write-back to lower memory hierarchy. So the SDC is propagated to the lower memory hierarchy. }
\end{itemize}

\subsubsection{{MESI Coherent Cache Line States}}
{In the MESI coherent cache architecture, we have the following cache status:}
\begin{itemize}
	\item
	{MESI\_INVALID: This state behaves the same as LINE\_NOT\_VALID state in the shared cache.}
	\item
	{MESI\_MODIFIED: This state behaves the same as LINE\_MODIFIED state in the shared cache.} 
	\item
	{MESI\_EXCLUSIVE: The cache line is only present in the current coherent cache and is not modified (hence, it is consistent with the lower memory hierarchy). Cache insert (evicting the old line) in this case does not need write-back to lower memory hierarchy. So the SDC is not propagated to the 
		lower memory hierarchy. In terms of SDC propagation, this state behaves the same as LINE\_VALID state in the shared cache.}
	\item
	{MESI\_SHARED: The cache line may be also stored in other caches of the machine and is clean (and is consistent with the lower memory hierarchy). Cache insert (evicting the old line) in this case does not need write-back to lower memory hierarchy. So the SDC is not propagated to the lower memory hierarchy. In terms of SDC propagation, this state behaves the same as LINE\_VALID state in the shared cache.}
\end{itemize}

\vspace{-0.3cm}
\subsection{{DL incidences}}
\begin{figure}
	\centering
	
	\includegraphics[width=0.5\textwidth]{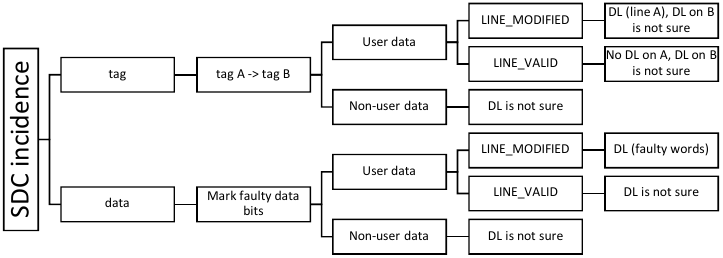}
	\vspace{-0.8cm}
	\caption{{Initial DL scenarios}}
	\label{fig:initial-dl}
	\vspace{-0.3cm}
\end{figure}

{We calculate the DL caused by an SDC in two phases: }
\begin{itemize}
	\item
	{Certain DL at the SDC incidence on user data.}
	\item
	{DL upon faulty cache access (DL propagation).}
\end{itemize}
{Fig.~\ref{fig:initial-dl} summarizes different scenarios for collecting DL statistics upon a SDC incidence.}

\subsubsection{{SDC on user data with LINE\_MODIFIED status}}
{When a SDC incidence occurs on user data, we count it as a cache line DL (SDC on tag) or word DL (SDC on data), if the affected line status is LINE\_MODIFIED (MESI\_MODIFIED in coherent cache). In the case of SDC on tag field, we flip the tag bits affected by SDC. Hence, valid tag A is changed to tag B while the line with tag B has all its data bits faulty (as the corresponding data belongs to another cache line with Tag A). If Tag B belongs to user data and the line status is LINE\_MODIFIED, we also have a cache line DL on address B, as line B will be written back to the lower memory hierarchy and will pollute its data. Please note that we do not simulate the cases the SDC changes the cache line status. }

\subsubsection{{SDC on user data with LINE\_VALID status}}
{If the SDC happens on a line with LINE\_VALID status, the DL incidence is not sure (as a valid copy of data is available in the lower memory hierarchy) and depends on later cache accesses (read/write/update/evict). In the case of fault on tag field, valid tag A is changed to tag B. Tag B, however, may refer to an invalid address space or an address space never used. There is also a possibility that tag B is accessed (read/write/update/evict). In this case, we are aware of the fact that this line (with tag B) has all its data bits faulty (as the corresponding data belongs to another cache line with tag A). }

\subsubsection{{SDC on non-user data}} 
\begin{figure}
	\centering 
	\includegraphics[width=0.42\textwidth]{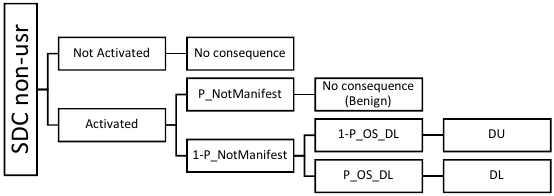}
	\vspace{-0.3cm}
	\caption{{Consequence of SDC on non-user data}}
	\label{fig:sdc-non-user-consequence}
	\vspace{-0.5cm}
\end{figure}

{Upon SDC on non-user data, the chance of DL is low but there is also a possibility that OS/applications malfunction result in DL in end-user data. 
	We can summarize the consequence of SDC on non-user data as shown in Fig.~\ref{fig:sdc-non-user-consequence}.
	Activated SDCs (SDCs that are accessed within execution) on non-user data lines are not manifested by the probability of $P_{NotManifest}$. In this case, the SDC has no visible abnormal impact in the system level (Benign SDC). 
	With the probability of $1-P_{NotManifest}$, the SDC is manifested (malign SDC) and has DU/DL consequences. 
	Malign SDCs result in DL (end-user data loss) by the probability of $P_{OS_{DL}}$ and result in DU by the probability of $1-P_{OS_{DL}}$. However, there are many cases that SDC is masked before a read access, as detailed in Fig.~\ref{fig:cache-access-nuser}.}

\vspace{-0.3cm}
\subsection{{Tracking DL propagation}}
\begin{figure}
	\centering 
	\includegraphics[width=0.5\textwidth]{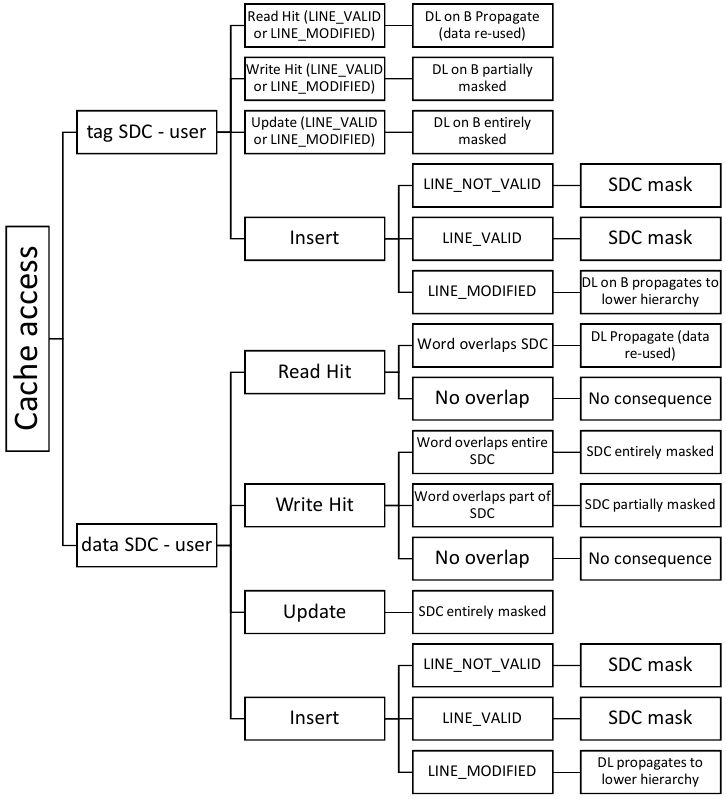}
	\vspace{-0.6cm}
	\caption{{Access SDC on user data}}
	\label{fig:cache-access-user}
	\vspace{-0.4cm}
\end{figure}

\begin{figure}
	\centering 
	\includegraphics[width=0.5\textwidth]{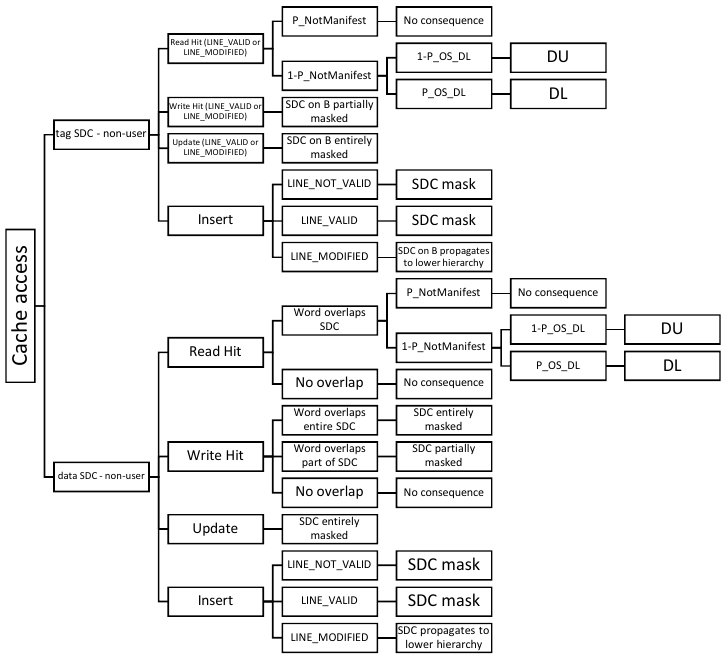}
	\vspace{-0.6cm}
	\caption{{Access SDC on non-user data}}
	\label{fig:cache-access-nuser}
\end{figure}

{The DL incidence and propagation, as well as fault masking, is checked upon each access (read, write, update, and evict) to the faulty cache line. Fig.~\ref{fig:cache-access-user} and Fig.~\ref{fig:cache-access-nuser} respectively show the scenarios of accessing SDC on user data and non-user data.}

\subsubsection{{Read}} 
{Reads a memory word. In this case the faulty data is used and the DL is propagated. We keep record of the times each SDC is used and count it in our DL calculation. The SDC may also propagate (write-back on a cache evict) to the lower memory hierarchy and be later used there. The magnitude of data loss is evaluated regarding the times the DL is used. As such, reading a faulty word is counted as one word (8 Bytes) DL despite the SDC is on tag or data. Please note if the SDC is on the data field, the read word is faulty if it overlaps the SDC location. Moreover, in the case of tag SDC, the DL occurs just if address B belongs to user data. Otherwise, if address B belongs to non-user data (i.e., due to the soft-error, address A belonging to user data is changed to address B belonging to non-user data), it is behaved as SDC in non-user data.}

\subsubsection{{Write}} 
{Writes a memory word. In this case, if the SDC is on data field and the written word overlaps the SDC, it masks the SDC (or parts of SDC, in the case the SDC affects two adjacent words). If the SDC is on the tag field, the written word overwrites the previous faulty word (that belongs to cache line A), while the rest of not-overwritten words already remains faulty. We added a data structure to each cache line to track such cases and keep record of the overwritten words. Hence, on a cache evict, we calculate the magnitude of data loss as shown in Equation~\ref{equ:cache-evict-mdl}.}
\begin{equation}
	\label{equ:cache-evict-mdl}
	\begin{split}
		\resizebox{1\hsize}{!}{$Magnitude~of~DL~(Bytes) = Line~Size $~-~$ (number~of~overwritten~words \times Word~Size)$}
	\end{split}
\end{equation}

\subsubsection{{Update}}
{Overwrites the whole data field. In this case, the SDC on the data field is entirely masked. In the case of SDC on the tag field, the wrong data of line B (that belongs to another cache line A) is entirely overwritten by a new data. Hence the SDC is entirely masked. }

\subsubsection{Evict (Insert new line)} 
{Evicts the old cache line if modified (LINE\_MODIFIED or MESI\_MODIFIED status) and writes a new cache line (with new tag and data). This operation masks SDC on both tag and data fields. However, if the old line has LINE\_MODIFIED (or MESI\_MODIFIED) status, it needs a write-back to the lower memory hierarchy and SDC/DL is propagated. }

\vspace{-0.3cm}
\subsection{{DUE}}
\begin{figure}
	\centering 
	\includegraphics[width=0.5\textwidth]{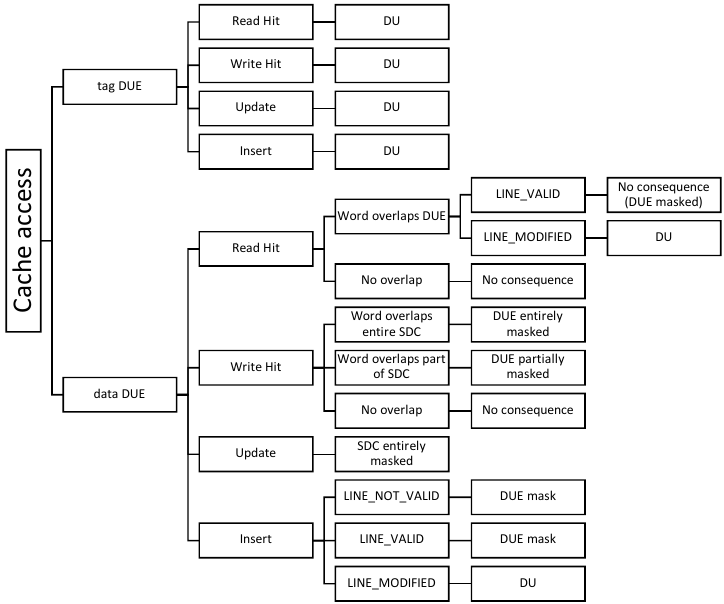}
	\vspace{-0.8cm}
	\caption{{Access DUE}}
	\label{fig:cache-access-due}
	\vspace{-0.5cm}
\end{figure}

{Upon DUEs on data field, in the case of reading a valid line (LINE\_VALID), the corrupted line is invalidated (LINE\_NOT\_VALID) by the cache controller and the correct data is obtained from lower memory hierarchies (hence, the fault is masked). In the case of reading a dirty line (with LINE\_MODIFIED status), however, the processor is rebooted (DU). 
	Upon cache write and update, as the data is overwritten and DUE is masked, the ECC is not checked. Upon cache evict (Insert), in the case cache line is not valid (LINE\_NOT\_VALID) or is clean (LINE\_VALID), the cache line is simply replaced with the new line. 
	However, in the case the evicted line is dirty (LINE\_MODIFIED), the evicted cache line needs to be written back to the lower memory hierarchy. 
	Hence, both tag and data ECCs are checked and in the case of error detection (DUE), the processor is rebooted (DU). 
	In the case of error on tag field, however, there is a possibility that the cache status is changed from LINE\_MODIFIED to LINE\_VALID or LINE\_NOT\_VALID. Hence, the cache controller takes the most conservative action and reboots the processor (resulting DU), regardless of the cache status. Fig.~\ref{fig:cache-access-due} summarizes the scenarios upon an access to a cache line with DUE.}

\vspace{-0.3cm}
\section{{SER Estimation of Cache Control Logic}}
\label{appendix:ser-control-logic}
\subsection{{Using Shivakumar et al.~\cite{shivakumar2002modeling} SER Data}}
{We obtain the SER for combinational logic, from the model by Shivakumar et. al.~\cite{shivakumar2002modeling}, which provides data for 600nm to 50nm technology nodes~\cite{shivakumar2002modeling}. Shivakumar et. al. report combinational logic SER for a logic chain of FO4 NAND gates (NAND4). For obtaining the total SER of a specific design, this method multiplies the number of logic to the SER per logic. The SER per logic chain of 2-input NAND4 logic is reported for the logic chains with the delay of 6, 8, 12, and 16 FO4 delay\footnote{{FO4 delay is the degree of pipelining characterized by the number of FO4 inverter gates that can be placed between two latches in a single pipeline stage~\cite{shivakumar2002modeling}. Hence, it is also representative of the working frequency.}} . FO4 delay is obtained by Equation~\ref{equ:fo4-delay}~\cite{sutherland1999logical}. }
\begin{equation}
	\label{equ:fo4-delay}
	\begin{split}
		FO4~Delay=360 \times L\_gate 
	\end{split}
\end{equation}

{Where $L\_gate$ is the transistor's drawn gate length in microns.} 

\subsubsection{{Obtain delay in terms of FO4 delay, as a function of working frequency}}

{FO4 delay for different technology nodes (180nm to 35nm) is reported in~\cite{ho2001future}. FO4 delay for 50nm technology node is 25ps~\cite{ho2001future}. Hence, the working frequency of a 50nm circuit is respectively 2.5 GHz, 3.3 GHz, 5 GHz, and 6.6 GHz for 16, 12, 8, and 6 FO4 delay.
	Assuming the frequency of 3.3 GHz for the target processor, the SER should be estimated by using 12 FO4 delay.}

\subsubsection{{Calculating the Length of Logic Chain}}

{The length of logic chain varies depending on the degree of pipelining characterized by the number of FO4 gates that can be placed between two latches in a single pipeline stage. Hence the exact length of logic chain (for 12 FO4) is $\frac{Inverter~Delay}{NAND~Delay}\times 12$.
	For estimating the ratio of inverter delay to NAND delay, we simply use the effort delay.
	Accordingly, the length of logic chain for 12 FO4 delay is 9 NAND gates.}

\vspace{-0.3cm}
\subsection{{Modeling Soft-Error in Cache Control Logic}}
{For modeling soft-error in cache control logic, we evaluate the area of control logic and covert it to the number of logic chains. Afterwards, we obtain SER per control logic. We assume each fault in control logic is manifested as a random change in cache tag address (we do not consider the effect of logical masking, as Shivakumar et al. do not). Please note as the error is initiated in the control logic, we assume it is not detectable/correctable by the cache ECC. Hence, each fault in the control logic is interpreted as an undetectable tag fault.} 

\subsubsection{{Evaluate Cache Logic Area}}
{To evaluate cache logic area we used CACTI 7.0~\cite{cacti7} that supports small cache sizes. We choose the CACTI configuration to be compliant with our simulated cache architecture in MARSSx86. Table~\ref{tab:cacti7-results} shows the L1 and L2 cache area obtained by CACTI 7.0. 
	We assume by excluding the memory cell area, the rest of cache area belongs to control logic. Note that this is not a careful assumption, as it ignores the area of components such as interconnections and networks.}

\begin{table}
	\centering
	\caption{{L1 and L2 cache area obtained by CACTI 7.0}}
	\vspace{-0.8cm}
	\begin{center}
		\begin{adjustbox}{width=\textwidth/2,totalheight=\textheight,keepaspectratio}
			\begin{tabular}{| c | c | c |}
				\hline
				& L1 & L2  \\ \hline
				Data Array Area ($mm^2$) & 0.99 & 10.89 \\ \hline
				Area Efficiency (Memory cell area/Total area) & 28.5 \% & 62.1 \% \\ \hline
				Tag Array Area ($mm^2$) & 0.027 & 0.57 \\ \hline
				Area Efficiency (Memory cell area/Total area) & 67.9 \% & 67.5 \% \\ \hline
				
			\end{tabular}
		\end{adjustbox}
	\end{center}
	
	\label{tab:cacti7-results}
	\vspace{-0.5cm}
\end{table}

\subsubsection{{Estimate the Control Logic Size in Number of NAND Gates}}
{We assume using standard 4-Transistor (4T) NAND gates in the control logic. To be compliant with CACTI area results obtained for 22nm technology node, the number of control logic transistors is estimated for Intel Haswell 22nm technology. To this end, we use total area and total number of transistors reported for Haswell family to estimate the number of transistors per area unit~\cite{haswell-processor}. 
	Finally, the number of 12 FO4 logics and SER of cache memory controller is shown in Table~\ref{tab:ser-cache-controller}.}

\section{{Impact of MBU Rates by Oliveira et al.~\cite{de2016evaluation}}}
{In this section, we investigate the MBU/MCU rates reported by Oliveira et al.~\cite{de2016evaluation}, summarized in Table~\ref{tab:mbu-oliveira}.
	Note that MBUs are defined as MCUs which target a single data word (hence, MBU incidences are also counted as MCU). 
	Oliveira et al. also mention that they have experienced up to 2-bit MBUs in their experiments~\cite{de2016evaluation}. 
	The details of applying empirical data of Oliveira et al. in our fault injection experiments is appeared in Appendix~\ref{appendix:mbu-oliveira}.
}

\begin{table}
	\centering
	\caption{{Percentage of faults (injected on Kepler GPU) that result in MBU/MCU~\cite{de2016evaluation}}}
	\vspace{-0.6cm}
	\begin{center}
		\begin{adjustbox}{width=\textwidth/2,totalheight=\textheight,keepaspectratio}
			\begin{tabular}{ | c | c | c | c | c | c | c |}
				\hline
				&	L2 0s (\%)	&L2 1s (\%)	&L2 Average (\%)	&L1 0s (\%)	&L1 1s (\%)	&L1 Average (\%)  \\ \hline
				1-bit	&67	&69	&68	&72	&74	&73 \\ \hline
				MCU	&33	&31	&32	&28	&26	&27 \\ \hline
				MBU	&6	&6	&6	&8	&7	&7.5 \\ \hline

			\end{tabular}
		\end{adjustbox}
	\end{center}
	
	\label{tab:mbu-oliveira}
\end{table}

\begin{figure}
	\begin{centering}
		\includegraphics[width=0.5\textwidth]{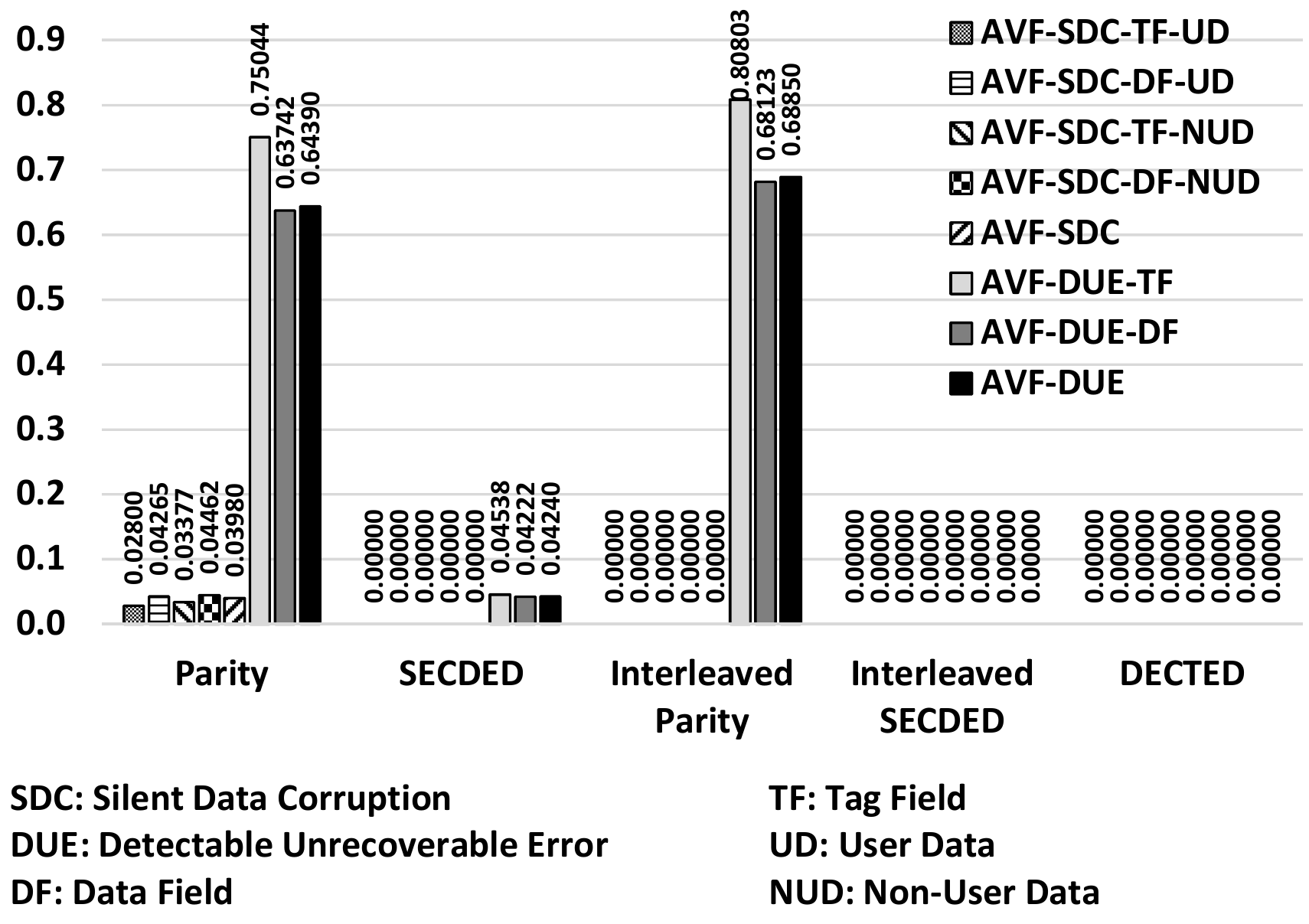}
		\caption{AVF for Different Cache Error Protection Schemes (Financial\_1 workload), using MBU rates by~\cite{de2016evaluation}. }
		\label{fig:protections-oliveira}
		\par\end{centering}
\end{figure}

{Fig.~\ref{fig:protections-oliveira} shows AVF values obtained for Financial\_1 workload using MBU/MCU {statistics} obtained by Oliveira et al.~\cite{de2016evaluation}. The figure shows that interleaved SECDED and DECTED protection schemes have all their AVF values equal to zero, as they both can correct up to two bit flips (the maximum magnitude of MBUs reported by Oliveira et al. is 2-bit~\cite{de2016evaluation}). The $AVF^{SDC}$ value is zero for all cache {protections}, except of parity code. The reason is that parity fails to detect 2-bit MBUs, while the rest of protections {can} either detect or correct 2-bit MBUs. We {also observe} that interleaved parity has slightly greater $AVF^{DUE}$ than parity code. The reason behind is the greater detection capability of interleaved parity, as the errors left undetected in parity code (resulting SDC) are {detected by interleaved parity, increasing $AVF^{DUE}$}.    
}

\begin{figure}
	\centering
	
	\includegraphics[width=0.5\textwidth]{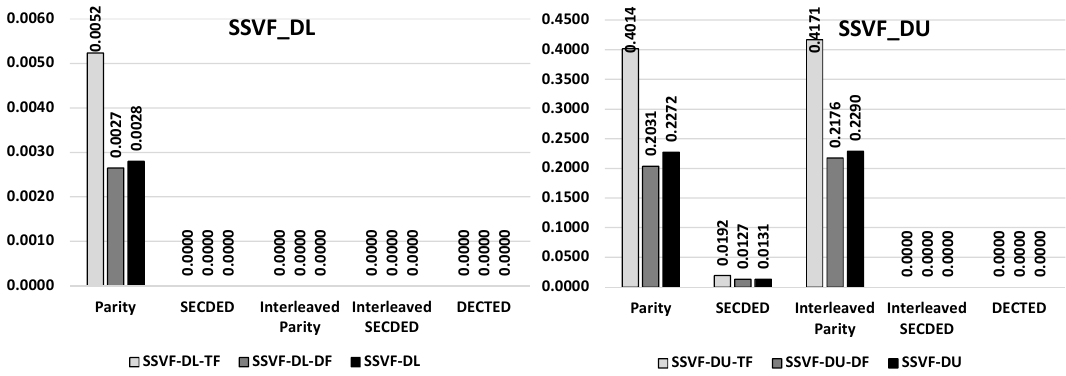}
	
	\caption{$SSVF^{DU}$ and $SSVF^{DL}$ for different cache protection schemes (Financial\_1 workload), using MBU rates by~\cite{de2016evaluation}.}
	\label{fig:svfdudl-oliveira}
\end{figure}

\begin{figure}
	\centering
	
	\includegraphics[width=0.5\textwidth]{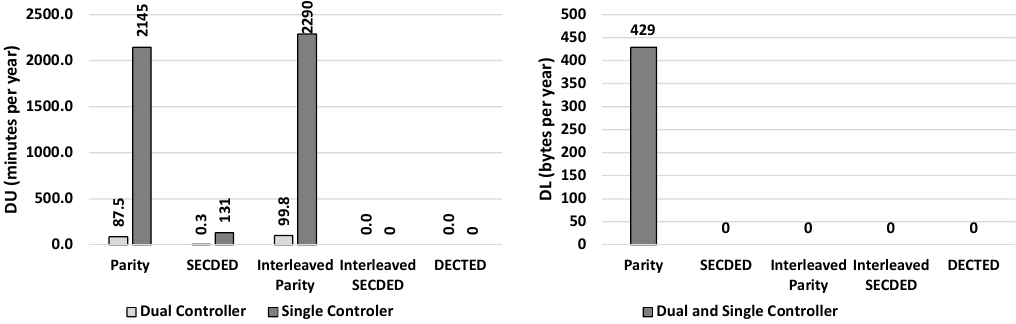}

	\caption{DU (minutes) and DL (bytes) in one year mission (assuming 1000 SEU/year) for single and dual controller with different cache protection schemes (Financial\_1 workload), using MBU rates by~\cite{de2016evaluation}.}
	\label{fig:normalizeddudl-oliveira}
\end{figure}

{Fig.~\ref{fig:svfdudl-oliveira} shows $SSVF^{DU}$ and $SSVF^{DL}$ {for} different cache protection schemes. As was expected from AVF values, {parity is} the only cache protection having non-zero $SSVF^{DL}$, as the rest of protections can either correct or detect up to 2-bit MBUs {(hence, no chance of DL when using MBU/MCU statistics of Oliveira et al.~\cite{de2016evaluation})}.
	Fig.~\ref{fig:normalizeddudl-oliveira} shows DU and DL for Financial\_1 workload for different cache protection mechanisms. The only cache protection that faces DL within mission time is parity code, as was also expected from its $SSVF^{DL}$ value. About DU, both interleaved SECDED and DECTED protections {experience} zero DU, as they both can correct up to 2-bit MBUs.
}

\begin{figure}
	\centering
	\subfigure[Share of MBU in total DU/DL for different cache protection schemes] 
	{
		\includegraphics[width=0.5\textwidth]{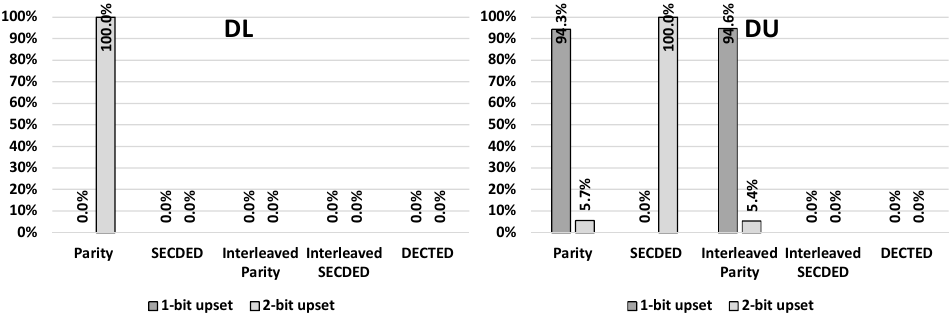}
		\label{fig:mbududl-oliveira}
	} 
	
	\subfigure[Number of DU and DL incidences by different type of MBU, normalized to the number of injected faults from each type of MBU] 
	{
		\includegraphics[width=0.5\textwidth]{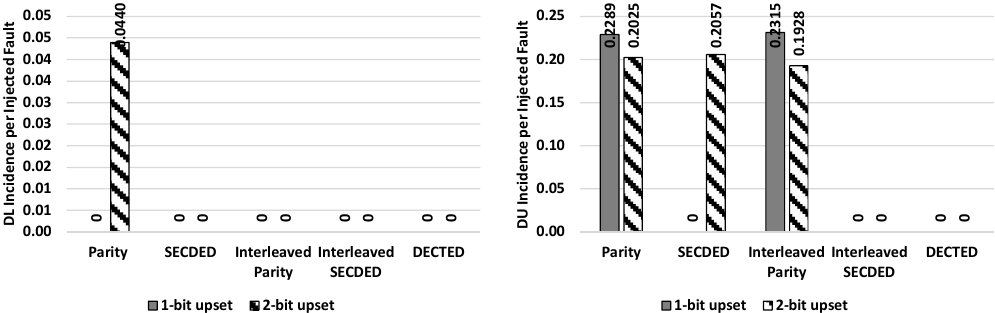}
		\label{fig:mbududl-normalized-oliveira}
	}

	\caption{Effect of different MBU types on DU/DL of data storage system (Financial\_1 workload), using MBU rates by~\cite{de2016evaluation}.}
	\label{fig:effect-mbududl-oliveira}
	
\end{figure}

{Fig.~\ref{fig:mbududl-oliveira} shows the share {of} MBUs in total DU/DL for different cache protection {schemes}. As the figure shows, more than 94\% of DU is caused by either 1-bit incidences or MCUs (that target more than one memory word). About DL, all is caused by 2-bit MBUs in the case of parity protection, while the rest of protection schemes experience no DL. 
	Fig.~\ref{fig:mbududl-normalized-oliveira} shows the number of DU and DL incidences by different types of MBU, normalized to the number of injected faults from each type of MBU. In DU {statistics}, the greater contribution of 1-bit compared to 2-bit is due to the fact that MCUs are counted as 2-bit MBU, but they flip {two single bits of two adjacent memory words} (hence, have a 1-bit upset effect {in each memory word}).}

\begin{figure}
	\centering
	
	\includegraphics[width=0.5\textwidth]{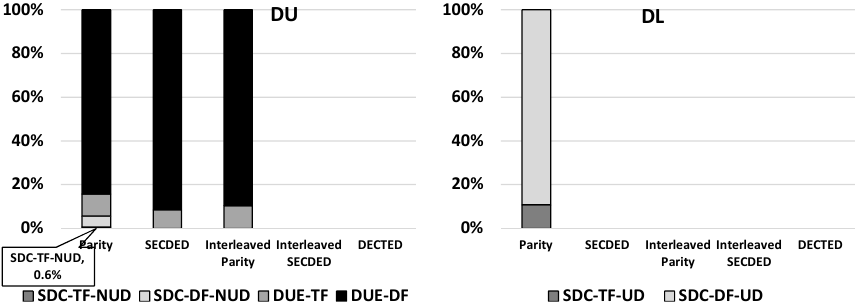}

	\caption{DU/DL breakdown for different cache protection schemes (Financial\_1 workload), using MBU rates by~\cite{de2016evaluation}.}
	\label{fig:dudlbreakdown-oliveira}
\end{figure}

{Finally, Fig.~\ref{fig:dudlbreakdown-oliveira} shows the contribution of different processor-level incidences in total DU/DL. As the figure shows, the most DU is caused by DUE on cache data field, while the most DL (in the case of parity code) is caused by SDCs on data field of user cache blocks.}

\vspace{-0.3cm}
\section{{Empirical Data of Oliveira et al.~\cite{de2016evaluation} used in Fault Injection Experiments}}
\label{appendix:mbu-oliveira}

{To apply the empirical data of Oliveira et al.~\cite{de2016evaluation} to our study, we modified our fault-injection engine to inject different rates of MBUs to L1 and L2 caches. Regarding {the} empirical results of Oliveira et al.~\cite{de2016evaluation}, we assume that MBUs have at most 2-bit size. We also assume that an MCU (that does not affect a single data word), pollutes two adjacent data words (the last bit of Line 0 Word 0 and the first bit of Line 0 Word 1). If the MCU happens in the last data word of a cache line, we assume it also pollutes the first word of the next cache line (Line 0 Word 7 and Line 1 Word 0). 
	Please note that Oliveira et al.~\cite{de2016evaluation} {do not} provide details on the spatial distribution of MCUs in the cache memory. We also have no information {about the spatial characteristics of} MCUs {when they target} the tag field (that highly depends on the physical alignment of tag/data in the cache architecture). So we simply assume that an MCU in the tag field targets two adjacent tags. In specific, we assume that a 2-bit MCU in tag field pollutes the tag of cache Set 1 Way 0, and Set 1 Way 1. If the MCU happens in the last way of a cache set, it pollutes the tag of Set 1 Way 7 and Set 2 Way 0. }

\end{document}